\documentclass[twocolumn, tighten, times]{aastex631}
\usepackage{graphicx,verbatim} 

\usepackage[T1]{fontenc}
\begin{document}
\title{Emission from multiple molecular isotopologues in a high-inclination protoplanetary disk}

\correspondingauthor{Colette Salyk}
\email{cosalyk@vassar.edu}

\author[0000-0003-3682-6632]{Colette Salyk}
\affiliation{Vassar College, 124 Raymond Avenue, Poughkeepsie, NY 12604, USA}

\author[0000-0001-7552-1562]{Klaus M. Pontoppidan}
\affiliation{Jet Propulsion Laboratory, California Institute of Technology, 4800 Oak Grove Drive, Pasadena, CA 91109, USA}
\affiliation{Division of Geological and Planetary Sciences, California Institute of Technology, MC 150-21, Pasadena, CA 91125, USA}

\author[0000-0003-4335-0900]{Andrea Banzatti}
\affiliation{Department of Physics, Texas State University, 749 N Comanche Street, San Marcos, TX 78666, USA}

\author[0000-0003-4179-6394]{Edwin Bergin}
\affiliation{Department of Astronomy, University of Michigan, 1085 S. University, Ann Arbor, MI 48109, USA}

\author[0000-0003-2631-5265]{Nicole Arulanantham}
\affiliation{Space Telescope Science Institute, 3700 San Martin Drive, Baltimore, MD 21218, USA}

\author[0000-0002-5758-150X]{Joan Najita}
\affiliation{NSF’s NOIRLab, 950 N. Cherry Avenue, Tucson, AZ 85719, USA}

\author[0000-0003-0787-1610]{Geoffrey A. Blake}
\affiliation{Division of Geological and Planetary Sciences, California Institute of Technology, MC 150-21, Pasadena, CA 91125, USA}

\author[0000-0002-6695-3977]{John Carr}
\affiliation{Department of Astronomy, University of Maryland, 4296 Stadium Dr., College Park, MD 20742, USA}

\author[0000-0002-0661-7517]{Ke Zhang}
\affiliation{Department of Astronomy, University of Wisconsin-Madison, Madison, WI 53706, USA}

\author[0000-0001-8184-5547]{Chengyan Xie}
\affiliation{Lunar and Planetary Laboratory, The University of Arizona, Tucson, AZ 85721, USA}

\begin{abstract}
We present a MIRI-MRS spectrum of the high-inclination protoplanetary disk around the solar-mass (K0) star MY Lup, obtained as part of the JWST Disk Infrared Spectral Chemistry Survey (JDISCS). The spectrum shows an unusually weak water emission spectrum for a disk around a star of its spectral type, but strong emission from CO$_2$, HCN, and isotopologues of both molecules.  This includes the first ever detection of C$^{18}$O$^{16}$O and H$^{13}$CN in an inner disk, as well as tentative detections of  C$^{17}$O$^{16}$O and HC$^{15}$N. Slab modeling provides molecular temperatures, column densities and emitting areas of the detected molecules. The emitting molecular gas is cold compared to that of other observed protoplanetary disk spectra. We estimate the isotopologue ratios of CO$_2$ and HCN, albeit with significant uncertainty. We suggest that the unusual spectrum of MY Lup arises from a combination of inner disk clearing, which removes emission from warm water, and its nearly edge-on inclination, which enhances line-of-sight column densities, although unusual chemistry may also be required.  MY Lup's spectrum highlights the potential to detect and measure trace isotopologues to study isotopic fractionation in protoplanetary disks; observations at higher spectral resolving power is needed to constrain the isotopologue ratios to greater precision.
\end{abstract}

\section{Introduction}

The observational study of protoplanetary disk chemistry in the few AU region was set in motion by the discovery of a forest of molecular emission lines \citep{Carr08,Salyk08} using the Spitzer InfraRed Spectrograph (Spitzer-IRS; \citealp{Houck04}).  That work detected and characterized the emitting properties of several simple molecules including water, HCN, C$_2$H$_2$, CO$_2$ and OH \citep{Carr11}, and allowed for basic accounting of key atomic species O,C and N in the inner disk \citep{Pontoppidan10b}. 

However, {\it cosmochemistry} in a solar system context often depends on a close attention to isotopic ratios.  In the solar system, isotopic uniformity of many heavier refractory elements suggest substantial mixing prior to the formation of most planetary building blocks \cite[e.g.][]{Zhu01,Moynier09,Pringle13}.  But small heterogeneities, especially of the volatile ``CHON'' atoms, tell a more complex story.  Each CHON atom may have a distinct isotopic history, including mixing between primordial reservoirs influenced by stellar nucleosynthesis and more processed solar-system materials, plus combinations of mass-dependent and mass-independent fractionation during early solar system evolution.  Isotope ratios are often distinct for certain classes of solar system objects, such that they can be used to trace back the origins of a given object's volatiles \citep[e.g.][]{Bockelee-Morvan15,Lis13}.   
 
Hydrogen, as measured primarily in water, shows enhancement in D/H in comets as compared to the protosolar value and some planetary materials, reflecting possible inheritance of materials deuterated in the cold interstellar medium (ISM; \citealp{Bockelee-Morvan15}). Earth also has a relatively high D/H ratio, suggesting a possible outer solar system contribution to its water reservoir \citep[e.g.][]{Bockelee-Morvan15}.  Nitrogen, as measured in CN and HCN, shows a rough trend of increased enhancement in heavier $^{15}$N from the protosolar value, to inner solar system bodies, and then to outer solar system bodies, like comets \citep{Marty12}.  Carbon shows a wide range of $^{13}$C/$^{12}$C ratios as measured in presolar grains, likely reflecting pollution by extraterrestrial materials influenced by nucleosynthesic processes \citep{Karhu23}.  Other solar system materials show much smaller heterogeneities, reflecting some enhancement of $^{13}$C relative to the bulk solar value  \citep{Karhu23}.  
 
Oxygen isotope ratios lie on a distinct slope-1 mass-independent fractionation line in a three-isotope ($^{17}$O/$^{16}$O vs. $^{18}$O/$^{16}$O) plot \citep[e.g.,][]{Clayton73,McKeegan11}.  This may be caused by enhanced photodissociation of C$^{17}$O and C$^{18}$O relative to C$^{16}$O, due to C$^{16}$O self-shielding, followed by subsequent incorporation of $^{17}$O and $^{18}$O into water ice \citep{Thiemens83,Yurimoto04,Lyons05}.

%What are isotopic ratios in ISM?
%Isotopic ratios have been measured in the ISM, showing variations depending on galactocentric distance.
 It is not yet clear to what extent the complex isotopic histories of solar nebula materials translate to other protoplanetary disks, but models and observations of isotopologues in outer disks provide some insight. Disk models suggest that there should be a range of isotopic ratios varying as a function of both disk location and the nature of the specific molecules, due to such factors as the local UV radiation field, the effectiveness of grain shielding, and the extent to which a formation pathway for a given molecule comes from the isotopically ``heavy'' or ``light'' products of the dissociation process \citep[e.g.][]{Miotello14}.  Nitrogen isotopes observed with ALMA in HCN and CN show evidence that CN may arise from reservoirs unaffected by isotope-selective photodissociation \citep{Hily-Blant17}, while HCN may, in some parts of the disk, have isotopic enhancements due to selective photodissociation \citep{Hily-Blant19}.  Similarly, ALMA observations of C$_2$H and CO suggest that the two molecules arise from different isotopic reservoirs \citep{Bergin24}, and that the CO isotopic ratios are radius-dependent \citep{Yoshida22}.  Precise measurements of all oxygen isotopes in disks are challenging, but \citet{Smith09} used infrared absorption spectroscopy of an outer protoplanetary disk to measure three oxygen CO isotopologues and found enrichment of C$^{18}$O and C$^{17}$O, consistent with selective photodissociation. 
 
Observations of isotopic ratios in {\it inner} disks could be used to further test our fractionation hypotheses for the terrestrial planet forming-regions.  They could also confirm whether radial gradients exist that could be used to trace back the origins of exo-planetary volatiles.  However, since molecular carriers of rarer isotopes can be much less abundant than primary isotopologues, very high sensitivity infrared spectroscopy is needed to detect them.

 The James Webb Space Telescope's Mid-InfraRed Instrument \citep{Rieke15} Medium Resolution Spectrometer (hereafter JWST MIRI-MRS; \citealp{Wright23}) offers significant improvements in sensitivity to molecular emission compared to its predecessor, the Spitzer InfraRed Spectrograph \citep{Houck04}, primarily due to the telescope's aperture size, but also enhanced by the improved resolving power.  Therefore, JWST potentially offers a powerful new tool to identify and study less abundant molecules, including isotopologues \citep[e.g.][]{Grant23,Xie23}.  While MIRI-MRS does suffer from instrumental fringing, which can limit signal-to-noise ratios (SNR), this can be substantially mitigated by using an asteroid as a ``fringing calibrator'' \citep{Pontoppidan24}.
 
 However, detection of infrared emission from rare species is still subject to a physical limitation: for protoplanetary disks observed in the infrared, the presence of dust limits the observable column of gas to the disk upper atmosphere.  At least three conditions can lead to more favorable conditions to detecting isotopologues in emission.  In a first, overall enhancement of a given molecule, for example due to radial transport of icy pebbles followed by sublimation \citep{Bosman17}, will enhance the secondary/primary isotopologue line ratio as the molecular column density grows.  In a second, rare species may be revealed if dust opacities are reduced due to physical clearing or grain growth, essentially increasing the gas column above the $\tau=1$ surface.  Such a scenario has been suggested by \citet{Grant23}, and further explored by \citet{Vlasblom24} to explain a detection of $^{13}$CO$_2$ in the GW Lup disk.  In a third possibility, presented in this work, a favorable geometry could enhance the observable gas column by allowing the observer to look through a less dusty upper part of the disk atmosphere.  

In this work, we present MIRI-MRS observations of isotopologues of both CO$_2$ and HCN in the disk surrounding the young star MY Lup.   MY Lup is a K0, solar-mass ($\log M_*=0.09^{+0.03}_{-0.12}$) star \citep{Alcala17,Andrews18} with a disk inclination measured to be 73$\pm0$.1 degrees at millimeter (mm) wavelengths \citep{Huang18}, or 77$\pm1.5$ in the near-infrared \citep{Avenhaus18}.  MY Lup's Kepler lightcurve has characterized it as a so-called ``dipper'', with shallow, frequent, semi-periodic dips in brightness likely due to occultation by the high-inclination disk \citep{Bredall20}. Its disk shows possible evidence for inner clearing based on both infrared photometry \citep{Romero12} and mm-wave continuum imaging \citep{vanderMarel18}, although obscuration of the inner disk and star by the outer disk complicates interpretations of the interferometric data \citep{Alcala19}.  Its accretion rate of 10$^{-8}\,M_\odot \mathrm{yr}^{-1}$ may be similar to that of other full disks of comparable mass \citep{Romero12,Alcala19}, although accretion rates below $2\times10^{-10}\,M_\odot \mathrm{yr}^{-1}$ were also reported by \citet{Alcala17}.  This source was also observed by the Atacama Large Millimeter/Submillimeter Array (ALMA) at high resolution as part of the The Disk Substructures at High Angular Resolution Project (DSHARP; \citealp{Andrews18}).  The ALMA images show shallow but detectable dark bands at $\sim$50 and 190 AU, and bright rings at $\sim$130 and 260 AU \citep{Huang18}.  

We use the detected CO$_2$ and HCN isotopologues to determine the properties (temperature, column density and emitting area) of the emitting molecules, and also place constraints on the $^{12}$C/$^{13}$C, $^{16}$O/$^{18}$O and $^{16}$O/$^{17}$O ratios in these molecules.  In addition, we consider what scenario(s) could have caused this unique set of isotopologue detections in MY Lup.

\section{Data Acquisition and Reduction}
MY Lup was observed with MIRI-MRS as part of the Cycle 1 program 1584 (PIs: Salyk, Pontoppidan) on 2023 Aug 13.  It was observed in all four channels with a 4-point dither pattern optimized for all channels, with a total exposure time of 1668 s.   Raw data used in this paper can be found in MAST: \dataset[10.17909/rd5f-h484]{http://dx.doi.org/10.17909/rd5f-h484}.  MY Lup was reduced using the standard procedure developed by the JWST Disk Infrared Spectral Chemistry Survey (JDISCS) team, a procedure fully described in \citet{Pontoppidan24}. The reduction procedure utilizes the standard MRS pipeline \citep{Bushouse24} up to stage 2b, which produces three-dimensional cubes for every exposure.  For the data presented here, we used pipeline version 
1.15.0 and Calibration Reference Data System context 1253.  Our reduction procedure then deviates from the standard pipeline.  In short, a set of wavelength-summed 2D images are created for each channel and sub-band to locate the source, and then 1D spectra are extracted with a wavelength-dependent aperture size.  Each 1D spectrum is then divided by an ``empirical spectral response function" derived from the fringe calibrator asteroid, and then combined together.  Temporal changes in the fringing pattern imply that observations close in time usually yield the best fringe removal.  For this set of observations, we used the asteroid 526 Jena, observed on 2023 Sep 21.
%For this set of observations, we used the asteroid 515 Athalia, observed on 2023 Apr 15.

A continuum-subtracted spectrum was created following the procedure outlined in \citet{Pontoppidan24}\footnote{Continuum determination routine is available at https://github.com/pontoppi/ctool/}. It relies on the assumption that the continuum is produced by grains that emit blackbody emission plus broad solid-state features and, thus, that it can be approximated by the flux values between the much narrower gas-phase emission lines. To create the continuum spectrum, the original (line-rich) spectrum is first smoothed with a median filter of a specified box size (in this case, 41 pixels in width).  Points in the original spectrum with flux values a certain threshold below the smoothed spectrum (in this case below a factor of 0.998) comprise the estimated continuum, and a new spectrum is created by interpolating between these continuum points.  This procedure is repeated in an iterative fashion, in this case, for 10 iterations.  The final continuum is then further smoothed with a second order Savitzky-Golay filter \citep{Savitzky64} using the scipy \citep{Virtanen20} routine savgol\_filter and is then subtracted from the original spectrum to produce the emission spectrum discussed in the remainder of this work.

\section{Spectrum Basics}
\begin{figure*}[ht!]
\centering
\includegraphics[width=7.5in]{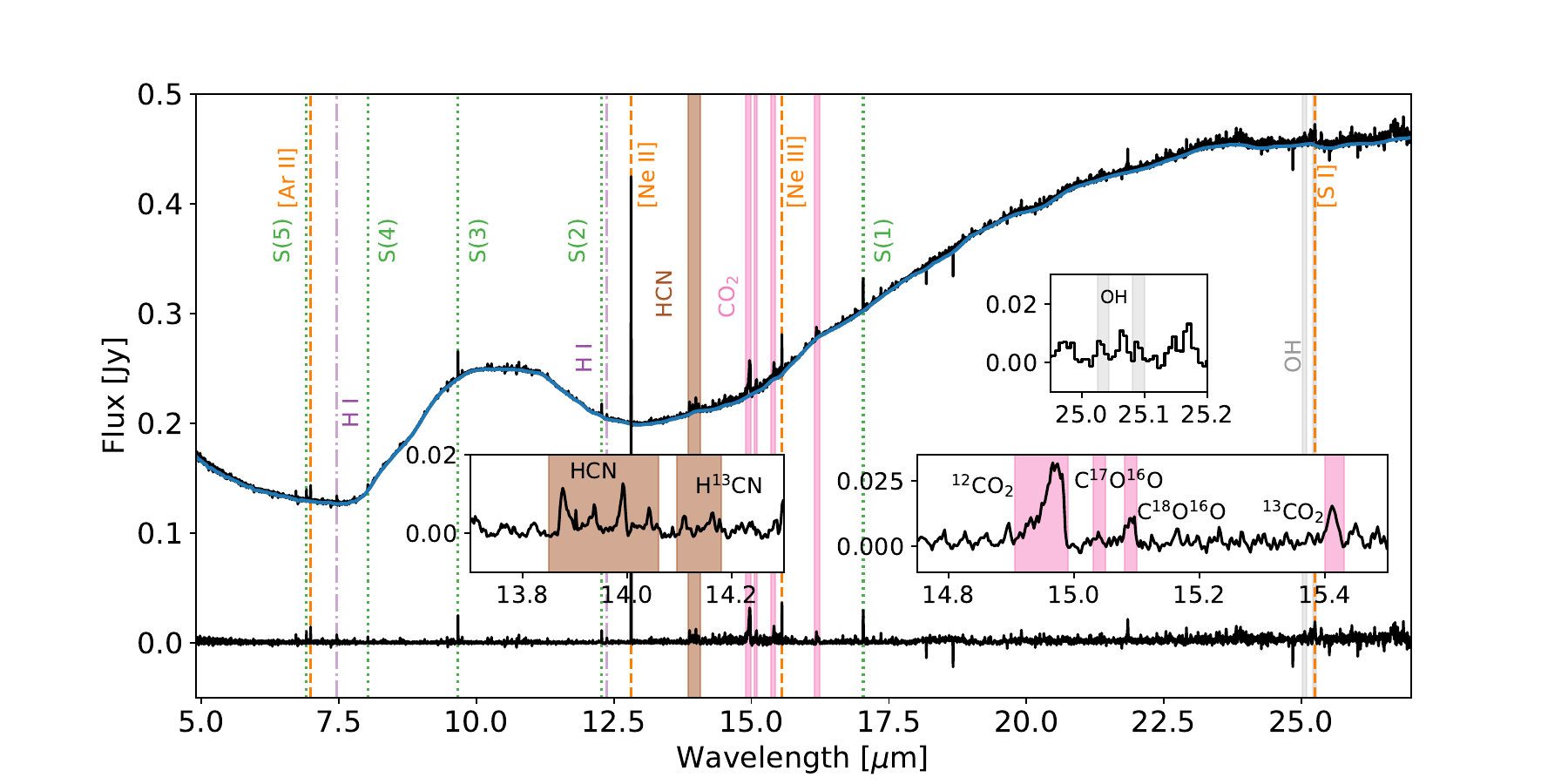}
\caption{Observed MIRI-MRS spectrum of MY Lup (black, top), continuum fit (blue) and continuum-subtracted flux (black, bottom). Prominent atomic and molecular emission features are labeled; S(X) refers to H$_2$ 0-0 transitions.  The broad feature at 10 $\mu$m is emission from solid silicates. Insets show continuum-subtracted spectra in the regions with significant HCN, CO$_2$ and OH emission; vertical bars also highlight locations of the main Q branches for different isotopologues.\label{fig:overview}}
\end{figure*}

The MIRI-MRS spectrum of MY Lup is shown in Figure \ref{fig:overview}, along with a continuum-subtracted spectrum.  The typical SNR is $\sim$300 based on line-free regions of the spectrum.  The spectrum shows prominent emission from ionized Neon and Argon, neutral Sulfur, molecular and atomic Hydrogen, as well as CO$_2$, HCN and OH. HCN, CO$_2$ and OH emitting regions are highlighted as insets.  Water vapor is also present in the spectrum, although features are less prominent than in other T Tauri disk spectra \citep[e.g.][]{Banzatti23}.  The unusual H I spectrum is explored further in Appendix \ref{sec:HI}.

\begin{figure*}[ht!]
\centering
\includegraphics[width=6.5in]{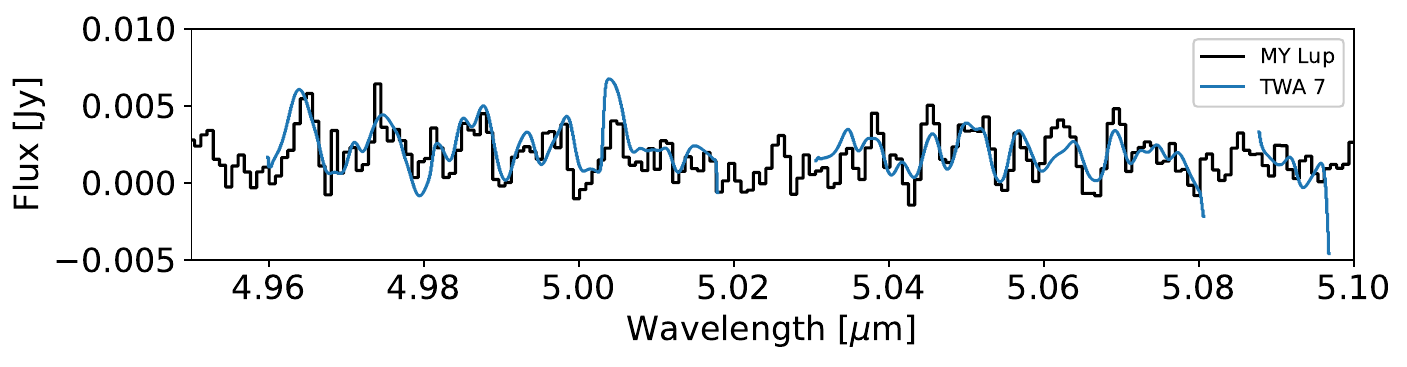}
\caption{Continuum-subtracted MIRI-MRS spectrum of MY Lup (black) and Keck-NIRSPEC spectrum of weak-line T Tauri star TWA 7, convolved to a resolving power of R=3000 (blue) and scaled to visually match the MY Lup spectrum.  Regions of the TWA 7 spectrum with high telluric contamination have been removed. \label{fig:photosphere}}
\end{figure*}

Unlike for MIRI-MRS spectra of other T Tauri disks \citep[e.g.][]{Temmink24a}, the spectrum shows no evidence for CO emission.  Figure \ref{fig:photosphere} shows instead that the $\sim$5 $\mu$m region has a spectral shape very similar to a spectrum of the weak-line T Tauri star TWA 7 --- the best apparent match from a small sample of weak-line T Tauri stars observed with Keck-NIRSPEC\footnote{Spectrum obtained by G. Blake and available at https://www.spexodisks.com/}.  This suggests that at $\sim$ 5 $\mu$m, the spectrum of MY Lup has detectable photospheric features and, therefore, low veiling (r, defined as the ratio of continuum to photospheric flux).  Note that veiling increases with wavelength as the stellar photospheric flux drops, so photospheric features become less apparent at longer wavelengths.  Based on the spectral scaling required to match the TWA 7 spectrum to the MY Lup spectrum, the M-band veiling would be r$_M\sim$ 4.5. However, TWA 7 has a spectral type of M2V, and is therefore a poor spectral type match to MY Lup.  An incorrect template will artificially increase the veiling value, so we consider the M-band veiling to be $<4.5$, in line with veiling values for known disks with inner clearings \citep{Salyk09}.  Low veiling has also recently been observed in the MIRI-MRS spectrum of the 30 Myr disk around WISE J044634.16–262756.1B, where it was also interpreted as evidence of inner disk dust depletion \citep{Long25}.  

\begin{figure*}[ht!]
\centering
\includegraphics[width=6.5in]{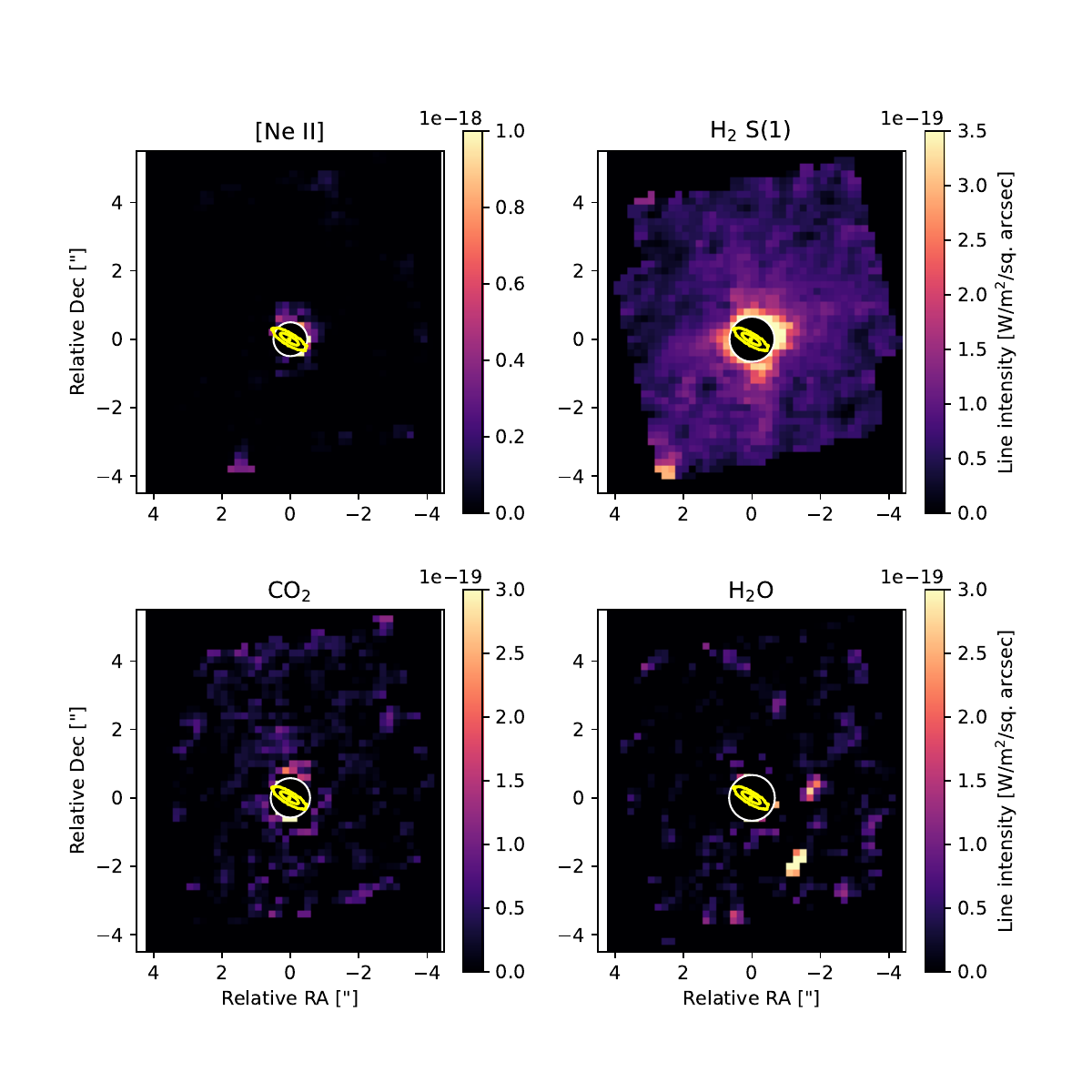}
\caption{Line images of [Ne II], H$_2$ S(1), $^{12}$CO$_2$ and a combination of 5 H$_2$O emission lines near 17 $\mu$m.  Circles mask an inner working angle of 1 x 1.22 $\lambda$/D around the star. Yellow contours show the ALMA continuum from \citet{Andrews18} (contours mark 5, 50 and 100$\sigma$). 
\label{fig:images}}
\end{figure*}

Spatial images are shown in Figure \ref{fig:images}.  Images are constructed by summing spectral channels near the line of interest, and then subtracting a continuum image constructed from adjacent channels on either side of the line.  The continuum image is also scaled to the same strength as the spectral sum, to isolate extended structures.  In the figure, we also mask an inner working angle of 1 x 1.22 $\lambda$/D around the star, and overplot the ALMA continuum from \citet{Andrews18} as contours.  As can be seen in Figure \ref{fig:images}, the H$_2$ S(1) line shows subtle evidence for a bipolar wide-angle wind perpendicular to the disk major axis; these images will be analyzed further in Pontoppidan et al. (2025, in preparation).  We see no evidence for spatially extended emission in [Ne II] nor in any of the molecular lines.  
%Huang18 PA = 58.8+/-0.1

\section{Analysis of Molecular Spectra}

\begin{deluxetable}{lcccccccc}
\tablehead{\colhead{Species}  &\colhead{N} & \colhead{T} & \colhead{R} & \colhead{Ratio\tablenotemark{a} } &\colhead{R$_i$/R$_\mathrm{ISM}$\tablenotemark{b}} \\
&\colhead{ [cm$^{-2}$]} &\colhead{[K]} & \colhead{[AU]} &  }
\startdata
\hline
H$_2$O  &9.9$\times 10^{17}$ &336 & $1.20$ \\  %R has been deprojected
\hline
$^{12}$CO$_2$ \tablenotemark{c}& 1.2$\times10^{17}$ & 575 & 0.39\\  %R has been deprojected
\hline
$^{12}$CO$_2$ &$3.5\times10^{18}$ & 325 & 0.58 & \nodata \\  %R has been deprojected
$^{13}$CO$_2$ & $5.1\times10^{16}$ & \textquotedbl & \textquotedbl & {\it 68}\tablenotemark{d} &1 \\
C$^{18}$O$^{16}$O & $1.8\times10^{16}$ & \textquotedbl & \textquotedbl & $381^{+132}_{-100}$ &$1.46^{0.52}_{-0.38}$\\
C$^{17}$O$^{16}$O & $3.1\times10^{15}$ & \textquotedbl & \textquotedbl &  $>2272_{-882}$ &$<0.88^{+0.56}$\\
\hline
$^{12}$CO$_2$ & $5.6\times10^{18}$ & 300 & 0.65& \nodata \\
$^{13}$CO$_2$ & $7.3\times10^{16}$ &\textquotedbl & \textquotedbl & $77^{+84}_{-25}$&$0.88^{+0.42}_{-0.46}$\\
C$^{18}$O$^{16}$O &$2.0\times10^{16}$ &\textquotedbl & \textquotedbl & 557 \tablenotemark{c}&1 \\
C$^{17}$O$^{16}$O &$3.6\times10^{15}$ &\textquotedbl & \textquotedbl & $>3109_{-1028}$ & $<0.64^{+0.32}$ \\
\hline
HCN & $1.4\times10^{19}$ & 250& 0.61\\
H$^{13}$CN & $2.0\times10^{17}$ & \textquotedbl & \textquotedbl & 68\\
\enddata
\tablenotetext{a}{{\it Atomic ratio} of the main light isotope divided by the selected heavy isotope.}
\tablenotetext{b}{Here, following cosmochemistry conventions, R$_i$ is the atomic ratio of the selected heavy isotope relative to the main isotope, i.e., the inverse of that used in the remainder of the text.  R$_i$/R$_\mathrm{ISM}>1$ implies enhancement of the {\it heavier} isotope relative to ISM values.}
\tablenotetext{c}{Fit including $^{12}$CO$_2$ only.  See discussion of degeneracies in Section \ref{sec:co2}.}
\tablenotetext{d}{Ratio fixed to the ISM value. }
\caption{\label{table:slab_fits} Summary of slab model fits}
\end{deluxetable}

\subsection{Water}
\label{sec:water}
\begin{figure*}[ht!]
\centering
\includegraphics[width=6.5in]{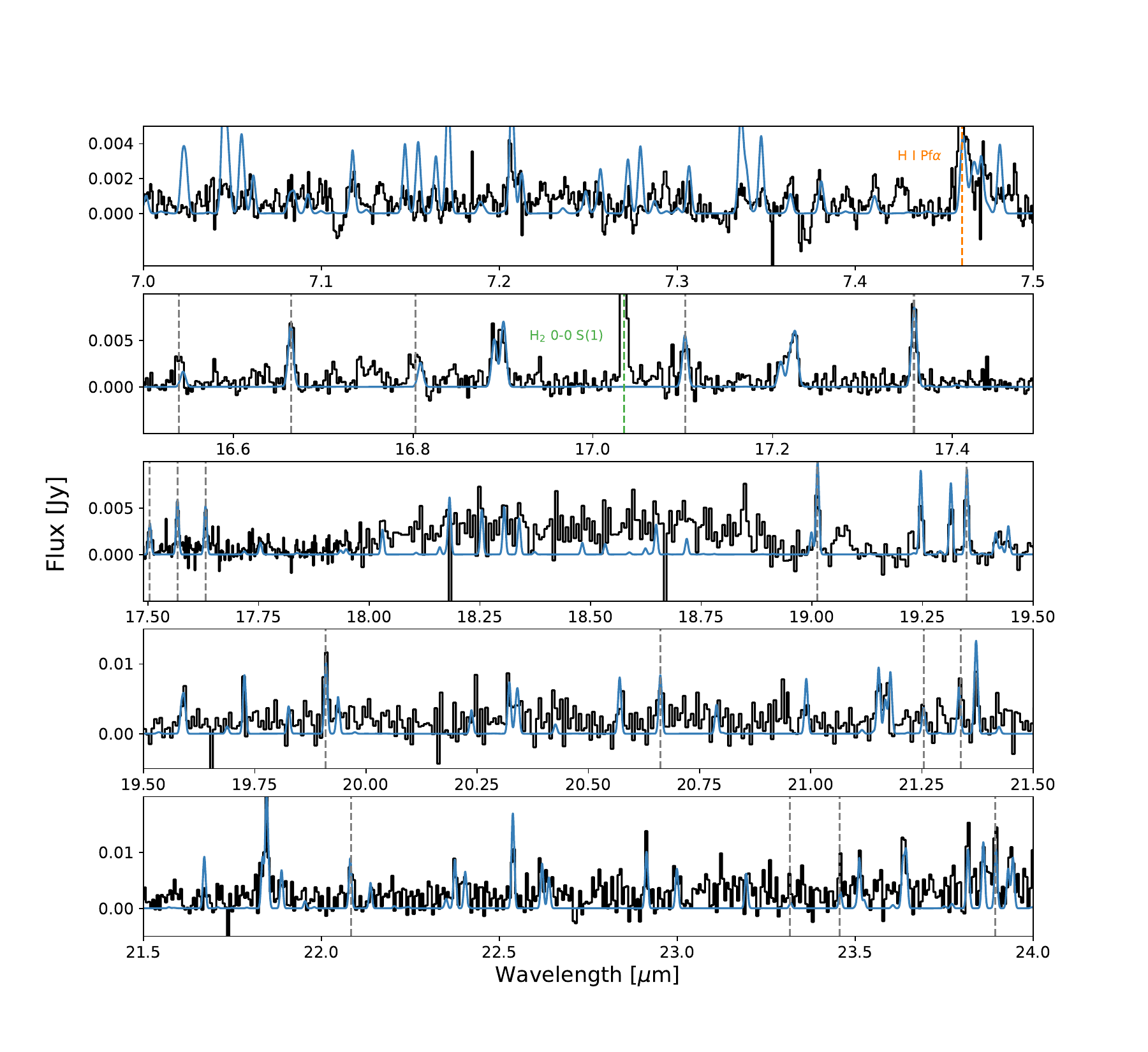}
\caption{Portions of the continuum-subtracted MY Lup MIRI-MRS spectrum (black) and slab water emission model (blue) with T=336 K, N=9.9$\times10^{17}$ cm$^{-2}$, R=1.20 AU.  Gray dashed lines mark locations of water lines used in the MCMC fit; note that rovibrational emission at 7 $\mu$ is not included in the fit.  H$_2$ and HI emission lines are also marked. \label{fig:h2o}}
\end{figure*}

Water vapor emission is less prominent in MY Lup's spectrum than in typical solar-like T Tauri stars \citep[e.g.][]{Pontoppidan10b}, but a few emission lines are observed.  Figure \ref{fig:h2o} shows the most prominent observed emission lines in the $\sim$17--24 $\mu$m regions, as well as the 7$\mu$m region, in which prominent bending mode rovibrational emission lines are seen in other targets \citep[e.g.][]{Banzatti23}.

To model the water emission, we use the HITRAN database \citep{Gordon22} to identify pure rotational water emission lines; we omitted rovibrational emission lines because their high critical densities likely result in non-Local Thermodynamic Equilibrium (non-LTE) excitation \citep{Meijerink09,Banzatti23}.  We then use the ``flux calculator'' routine in the spectools-ir python package \citep{Salyk22} to extract line fluxes.  We visually vet the lines, ultimately selecting fluxes from 18 isolated lines to fit, ranging from 17--24 $\mu$m with excitation temperatures between $\sim$1300 and 3600 K.  We then fit the emission line fluxes with the spectools-ir ``slab fitter'' routine; this package performs Markov Chain Monte Carlo fitting with the ``emcee'' package \citep{Foreman-Mackey13} using flat priors and a ``slab'' emission model, which treats the gas as a slab in LTE with a single temperature (T), column density (N, which need not be optically thin) and projected emitting area (A$_\mathrm{proj}$).  The projected emitting area can be converted to an emitting radius $R$ if we assume a disk-like geometry, i.e., A$_\mathrm{proj}=\pi R^2 \cos{i}$ where $i$ is the disk inclination.  Hereafter we take $i$ to be the 77$^\circ$ near-infrared inclination from \citet{Avenhaus18}.  Local line broadening is assumed to be thermal.  We set the limits of our priors to 150--1000 K for T, 14--20 for log (N $[\mathrm{cm}^{-2}])$ and $-4 - 4$ for $\log A_\mathrm{proj}$ [AU$^2$].  

The MY Lup water vapor emission is found to be consistent with T=336$\pm17$ K, log (N) = 18.00$\pm0.18$ (N=9.9$\times 10^{17}$ cm$^{-2}$) and $R=1.20\pm0.08$  AU (see also Table \ref{table:slab_fits}). The best-fit emission model is shown in Figure \ref{fig:h2o}; note that the rovibrational emission shown in the upper panel is not used in the fit due to the contribution of non-LTE excitation for these lines \citep{Banzatti24}.  

%R = R_proj / sqrt(cos i)

\begin{figure*}[ht!]
\centering
\includegraphics[width=6.5in]{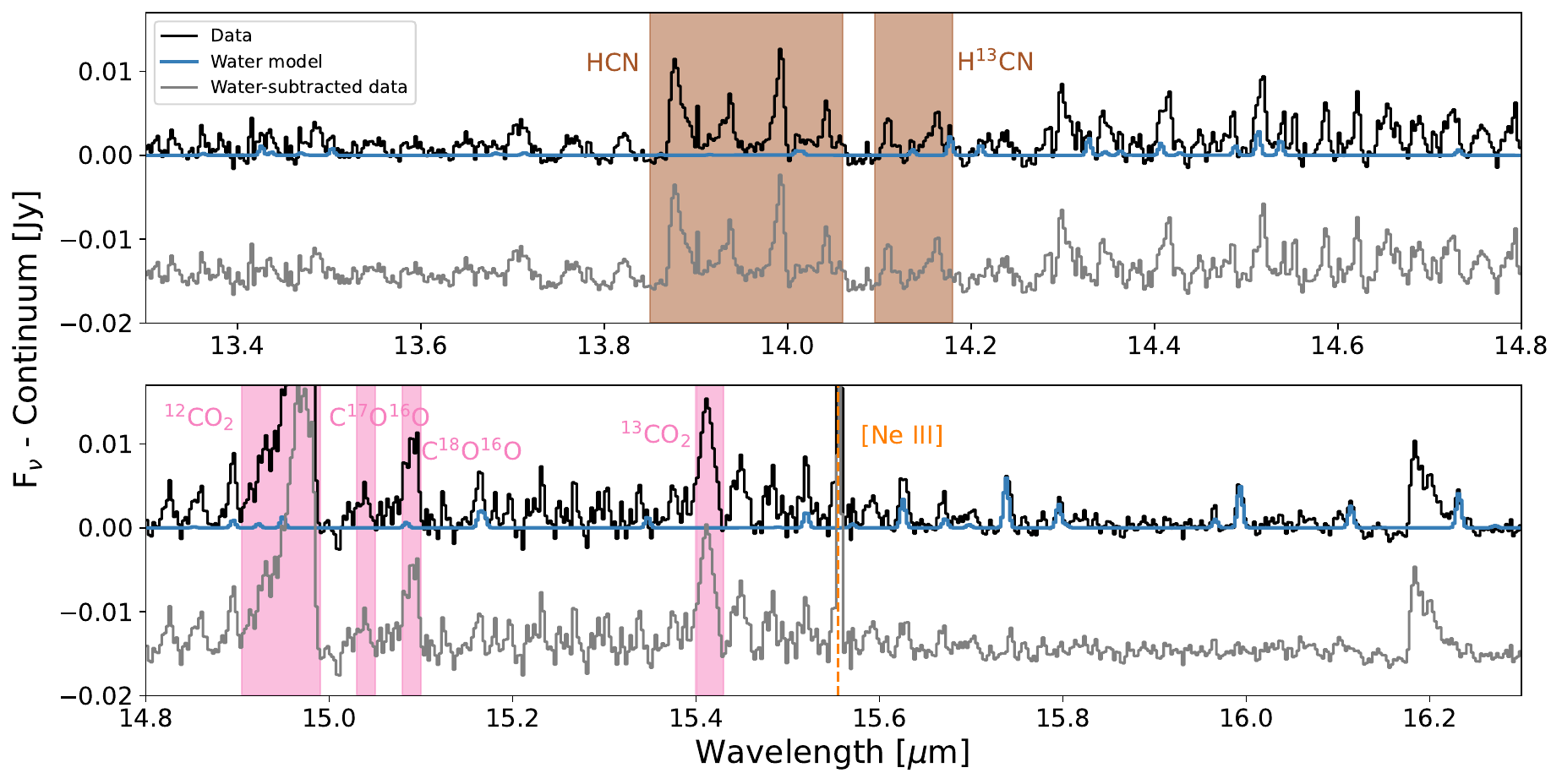}
\caption{Observed continuum-subtracted MY Lup spectrum (black), best-fit slab water model  convolved to a FWHM of 120 kms$^{-1}$ (blue), and residuals (gray, offset) in the HCN and CO$_2$-emitting regions.  Vertical bars mark the same molecular Q branches highlighted in Figure \ref{fig:overview}.
\label{fig:water_subtraction}}
\end{figure*}

For subsequent analysis in this work, this best-fit water model is subtracted from the data to produce our nominal water-subtracted spectrum.  In Figure \ref{fig:water_subtraction}, we show the impact of this water subtraction on the spectra in the HCN and CO$_2$ emitting regions.  We note that there is minimal water contamination in the main Q branches of HCN, CO$_2$ and their isotopologues.

\subsection{Carbon dioxide}
\label{sec:co2} 
\begin{figure*}[ht!]
\centering
\includegraphics[width=6.5in]{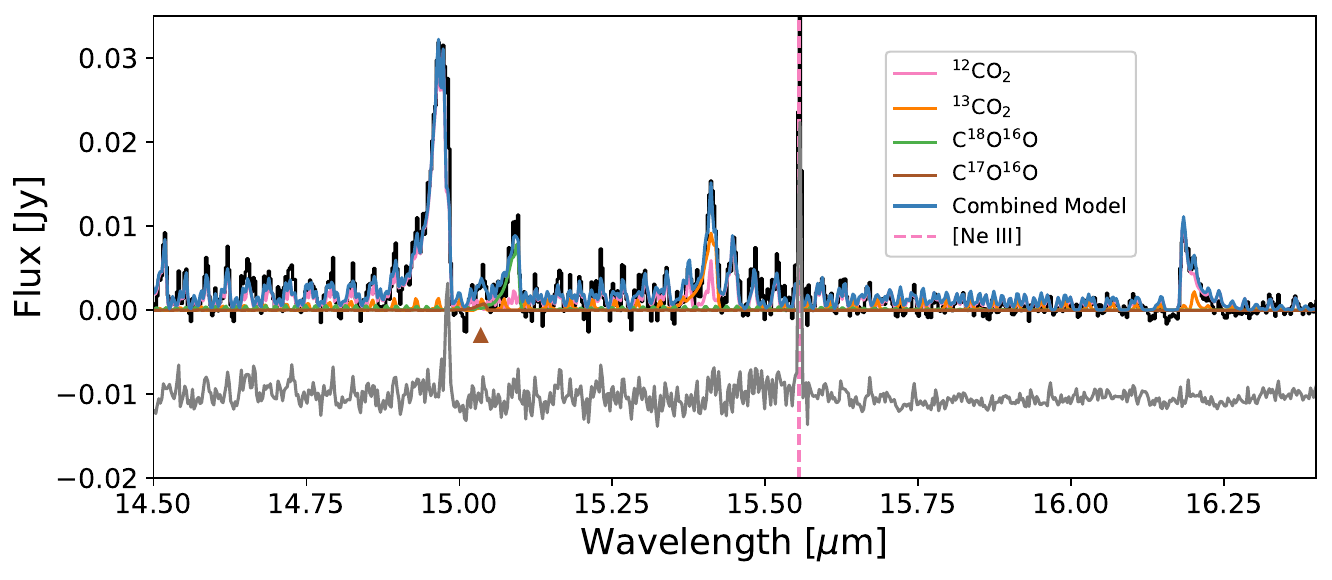}
\caption{MY Lup MIRI-MRS continuum-subtracted and water model-subtracted spectrum in the CO$_2$ emitting region (black) plus slab emission models for CO$_2$ and isotopologues (various colors) assuming a C ratio fixed to the ISM value --- see Table \ref{table:slab_fits} for model parameters.  The location of the marginal C$^{17}$O$^{16}$O detection is highlighted with a brown triangle. [Ne III] emission at 15.555 $\mu$m is also marked. Residuals (data - model) are shown in gray, offset by 0.01 Jy for clarity.   The large residual at $\sim$14.99 $\mu$m is discussed further in Appendix \ref{sec:nonLTE}.\label{fig:co2}}\end{figure*}

Figure \ref{fig:co2} shows the main CO$_2$ emitting region after subtraction of the best-fitting water model, along with CO$_2$ emission models, discussed below. The spectrum shows strong emission from CO$_2$, as well as the isotopologues $^{13}$CO$_2$ and C$^{18}$O$^{16}$O. There is also marginal evidence for  C$^{17}$O$^{16}$O, although this feature lies on top of a $^{12}$CO$_2$ P-branch line.   ([Ne III] also emits at 15.555 $\mu$m).

 We then fit the CO$_2$ emission spectrum with LTE slab models using the spectools-ir ``slabspec'' routine \citep{Salyk22}.  Unlike for the water emission, where isolated emission lines can be identified, and modeled and observed line {\it fluxes} can be compared, the CO$_2$ (and HCN) emission is highly blended, and a synthetic spectrum must be produced for comparison with the observed data.  Therefore, we explore parameter space with a less time-intensive grid of models (rather than MCMC).  We create a grid of slab CO$_2$ emission models with a range of column densities (from 15--20 for log N [cm$^{-2}$] with $\Delta$log N = 0.2) and temperatures (T from 200 K to 900 K with $\Delta$ T $= 25$ K).  For every CO$_2$ model, we adjust the model peak line fluxes to match the observed peaks, and then compute $\chi^2$. By adjusting the peak line flux values of the model, we are implicitly adding projected emitting area ($A_\mathrm{proj}$) as a third model parameter (see, e.g. \citealp{Grant23}).  $\chi^2$ is computed between 14.5 and 16.4 $\mu$m, excluding the $^{13}$CO$_2$, C$^{18}$O$^{16}$O, C$^{17}$O$^{16}$O and [Ne III] emitting regions.

\begin{figure*}[ht!]
\centering
\includegraphics[width=6.5in]{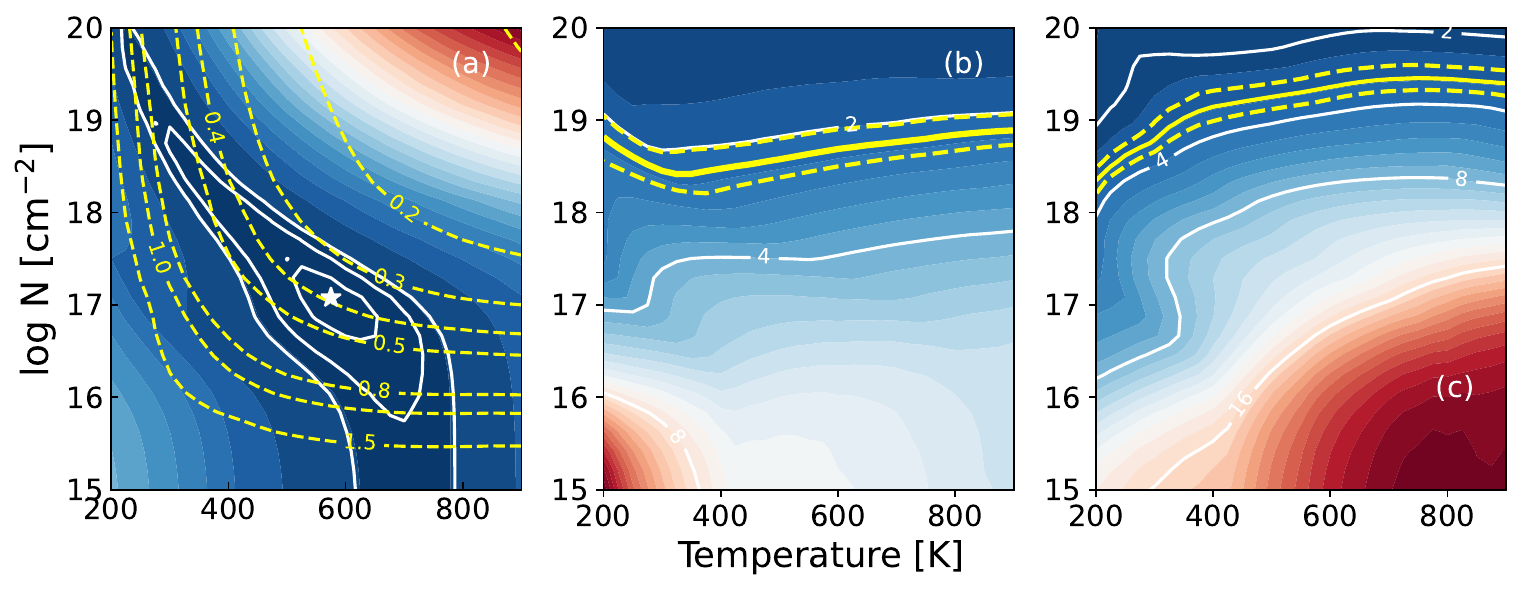}
\caption{(a) $\chi^2$ contour plot for fits to the CO$_2$ emission, as a function of temperature and CO$_2$ column density.  White solid contour lines mark 1,2, and 3-sigma confidence intervals on the best-fit model, shown with a white star.  Dashed yellow lines show best-fit emitting radii (R) corresponding to each model as yellow dashed lines. (b) Line peak ratio of the Q branches of $^{12}$CO and $^{13}$CO, assuming $^{12}C/^{13}C=68$.  Yellow lines show the  observed ratio of 2.24$\pm$0.24. (c) Line peak ratio of the Q branches of $^{12}$CO and C$^{18}$O$^{16}$O, assuming $^{12}$CO/C$^{18}$O$^{16}$O = 278.5.  Yellow lines show the observed ratio of 2.97$\pm$0.36.\label{fig:co2_ratios}}\end{figure*}

\begin{figure*}[ht!]
\centering
\includegraphics[width=6.5in]{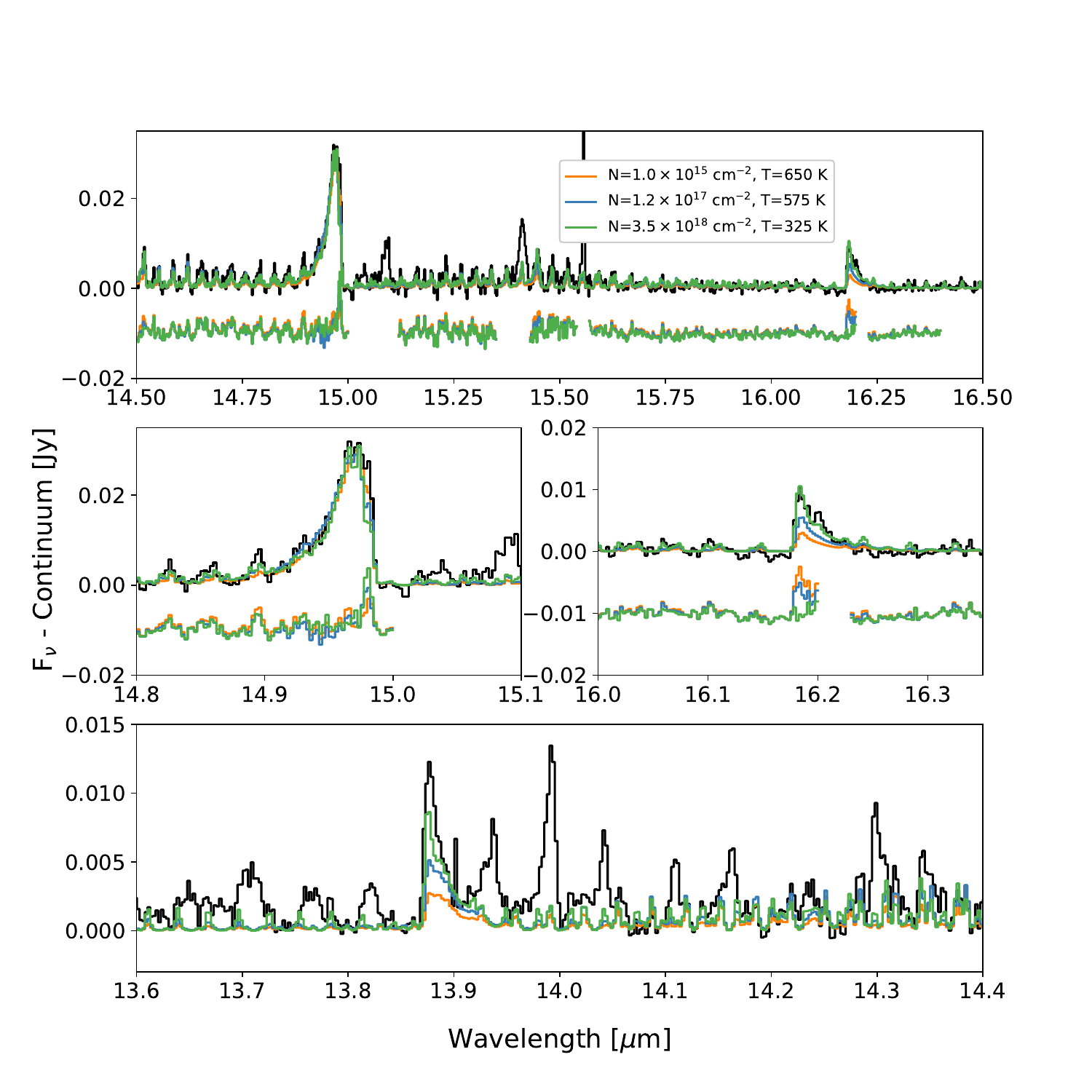}
\caption{Continuum and water-subtracted MIRI-MRS spectrum of MY Lup (black) compared with 3 slab models (colors).  Residuals (with regions contaminated by isotopologue or atomic emission removed) are shown below.  The middle panels highlight two regions with significant differences between the residuals. The bottom panel highlights a CO$_2$ feature in the HCN-emitting region, which was not included in the calculation of $\chi^2$. \label{fig:nonLTE}}
\end{figure*}

$\chi^2$ contours are shown in Figure \ref{fig:co2_ratios}(a).  White contours show 1,2, and 3 sigma confidence intervals on the models, assuming that the reduced $\chi^2=1$ for the best-fit model.  The best-fit model has T=575 K, N$=1.2\times10^{17}$ cm$^{-2}$ and A$_\mathrm{proj}$ = 0.11 AU$^{2}$ (R$ = 0.39$ AU; see Table \ref{table:slab_fits}).  However, there is a strong degeneracy between N and T.  A closer look at the model fitting and residuals is provided in Figure \ref{fig:nonLTE}.  We see that multiple models can match the data in many regions, with only a few features dominating the differences in residuals.  A close look at the residuals also shows that a feature on the red edge of the main CO$_2$ Q branch ($\sim$14.98 $\mu$m) prefers low N models, while a feature at $\sim$16.2 $\mu$m prefers high N models.  A $\sim$13.9 $\mu$m feature in the HCN-emitting region, which was excluded from the calculation of $\chi^2$ due to contamination from HCN, also prefers the higher N models.  The fact that different parts of the spectrum are consistent with different models may indicate contributions to the CO$_2$ spectrum from non-LTE excitation, which we discuss further in Appendix \ref{sec:nonLTE}.  Alternatively, the spectrum might be produced by disk regions with a range of temperatures and/or column densities.
 
%see MYLup_co2only_chisq.ipynb

A more stringent constraint on N can be provided by the relative strength of the Q branches for different isotopologues, {\it if} the abundance ratio of the two molecules is known and they are assumed to emit from the same reservoir.  Although slightly different temperatures were required to fit the $^{12}$CO$_2$ and $^{13}$CO$_2$ emission from GW Lup \citep{Grant23}, we find no clear evidence that different temperatures are required to fit the two different Q branches in MY Lup's spectrum.  In panel (b) of Figure \ref{fig:co2_ratios}, we show the line peak ratio of the $^{12}$CO$_2$ and $^{13}$CO$_2$ Q  branches assuming $^{12}$CO$_2$/$^{13}$CO$_2$ is set to the ISM $^{12}$C/$^{13}$C abundance ratio of 68 \citep[][See also Table \ref{table:ism_ratios}]{Milam05}.  The observed line peak ratio is  2.24$\pm$0.24, which implies $N_\mathrm{CO_2}\gtrsim10^{18}$ cm$^{-2}$ for this assumed molecular ratio.  In panel (c), we show the line peak ratio for the $^{12}$CO and C$^{18}$O$^{16}$O Q branches, assuming $^{12}$CO/C$^{18}$O$^{16}$O = 278.5 (which is half the ISM {\it atomic} ratio of 557; \citealp{Wilson99}, since there are two O's in CO$_2$).   The observed value of 2.97$\pm$0.36 is also consistent with $N_\mathrm{CO_2}\gtrsim10^{18}$ cm$^{-2}$.  The column density and the molecular ratio are degenerate, however; lower $^{12}$CO$_2$/$^{13}$CO$_2$ or $^{12}$CO/C$^{18}$O$^{16}$O ratios would imply a lower best-fit column density.

\begin{deluxetable}{lcl}
\tablehead{\colhead{Ratio}  & \colhead{Value} & \colhead{Reference} }
\startdata
$^{12}$C/$^{13}$C & 68 & \citet{Milam05}\\
$^{16}$O/$^{18}$O & 557 & \citet{Wilson99} \\
$^{16}$O/$^{17}$O & 2005&  \citet{Wilson99} \\
\enddata
\caption{\label{table:ism_ratios} ISM values of isotopic ratios assumed in this work}
\end{deluxetable}

\begin{figure*}[ht!]
\centering
\includegraphics[width=6.5in]{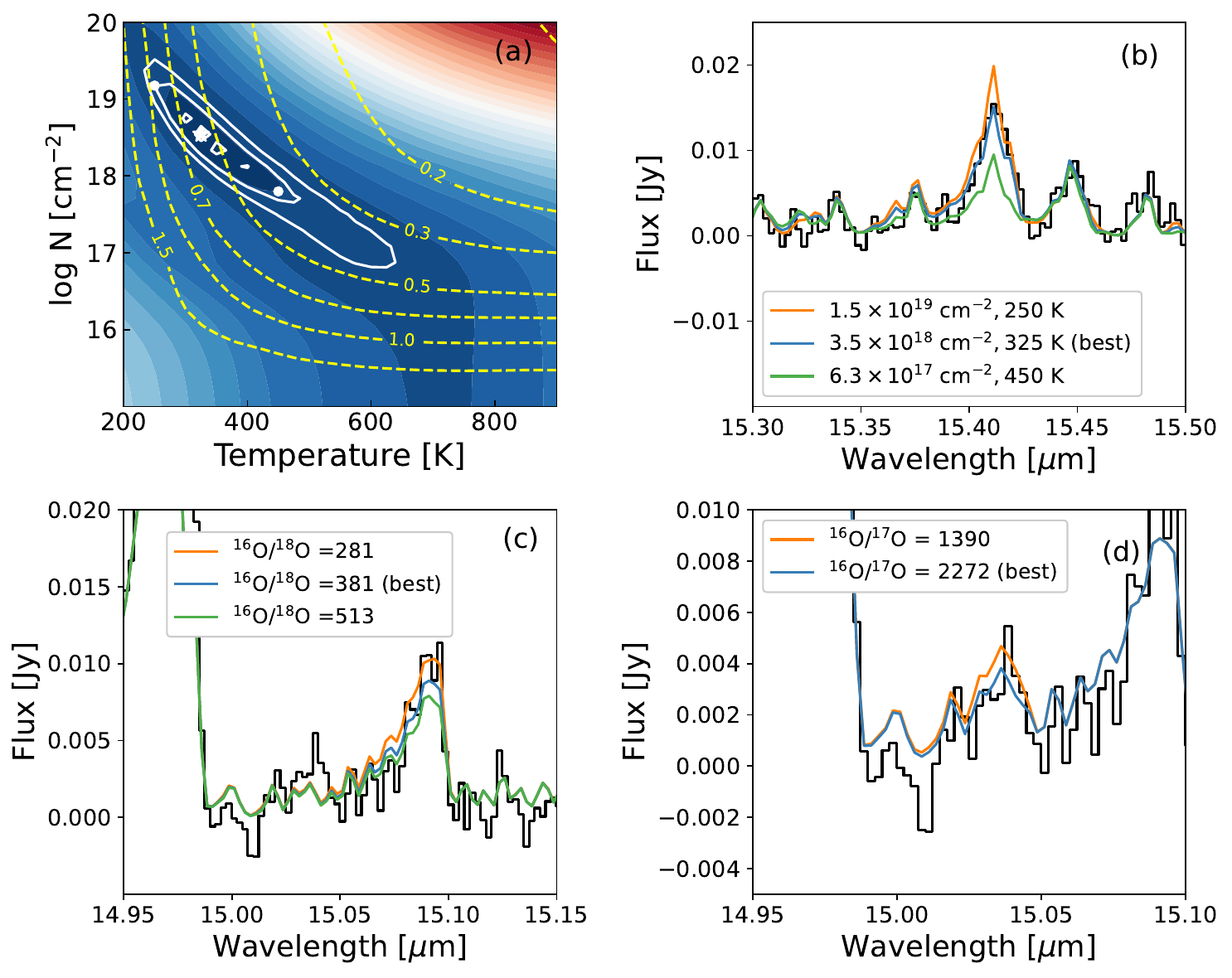}
\caption{Analysis with C ratios fixed to ISM values.  (a) $\chi^2$ diagram for combined $^{13}$CO$_2$ and $^{12}$CO$_2$ fit.   White star shows the best-fit model, while white circles show models highlighted in panel (b). White solid and yellow dashed lines have the same meaning as in Figure \ref{fig:co2_ratios}(a). (b) $^{13}$CO$_2$ Q branch overplotted with the best-fit model and two additional models at the edges of the 2$\sigma$ $\chi^2$ contour. Legend labels show the values of N$_\mathrm{CO_2}$ and T  for these models. (c) C$^{18}$O$^{16}$O Q branch overplotted with models assuming the best-fit $^{16}$O/$^{18}$O ratio and those representing the best fit $\pm1\sigma$.  (d) C$^{17}$O$^{16}$O Q branch overplotted with a model with the best-fit $^{16}$O/$^{17}$O as well as one representing the best fit $-1\sigma$.\label{fig:co2_fixedCratio}}
\end{figure*}

Given the degeneracies inherent in this modeling, we therefore cannot independently fit the column density and abundance ratio.  Instead, we test whether the ISM abundance ratios can provide a good fit to the data, and investigate the implications of the results.   We start by fixing the C ratios to the ISM value ($^{12}$C/$^{13}$C=68), fitting both the $^{12}$CO$_2$ and $^{13}$CO$_2$ Q branches, and using the resulting constraints to pinpoint the oxygen isotope ratios.  The $\chi^2$ diagram in Figure \ref{fig:co2_fixedCratio}(a) shows that using $^{13}$CO$_2$ along with $^{12}$CO$_2$ in the fit moves the best-fit model towards higher N$_\mathrm{CO_2}$ values (consistent with Figure \ref{fig:co2_ratios}(b)) and lower T values.  The best-fit model has N$_\mathrm{CO_2}=3.5\times10^{18}$ cm$^{-2}$, T$=325$ K and R$=0.58$ AU.  A degeneracy is still present due to the $^{12}$CO$_2$ issues highlighted in Figure \ref{fig:nonLTE}, but Figure \ref{fig:co2_fixedCratio}(b) shows that the $^{13}$CO$_2$ Q branch is very sensitive to N$_\mathrm{CO_2}$. 

Using the best-fit N$_\mathrm{CO_2}$ and T, we then vary the oxygen isotope ratios to find those most consistent with the data.  We find a best-fit $^{16}$O/$^{18}$O ratio of 381$^{+132}_{-100}$; the best fit and $\pm1\sigma$ models are shown in Figure \ref{fig:co2_fixedCratio}(c).  The C$^{17}$O$^{16}$O feature is consistent with a best-fit $^{16}$O/$^{17}$O ratio of 2272$_{-882}$, but higher ratios are also consistent with this marginal detection.  The best fit model and $-1\sigma$ model are shown in Figure \ref{fig:co2_fixedCratio}(d).

\begin{figure*}[ht!]
\centering
\includegraphics[width=6.5in]{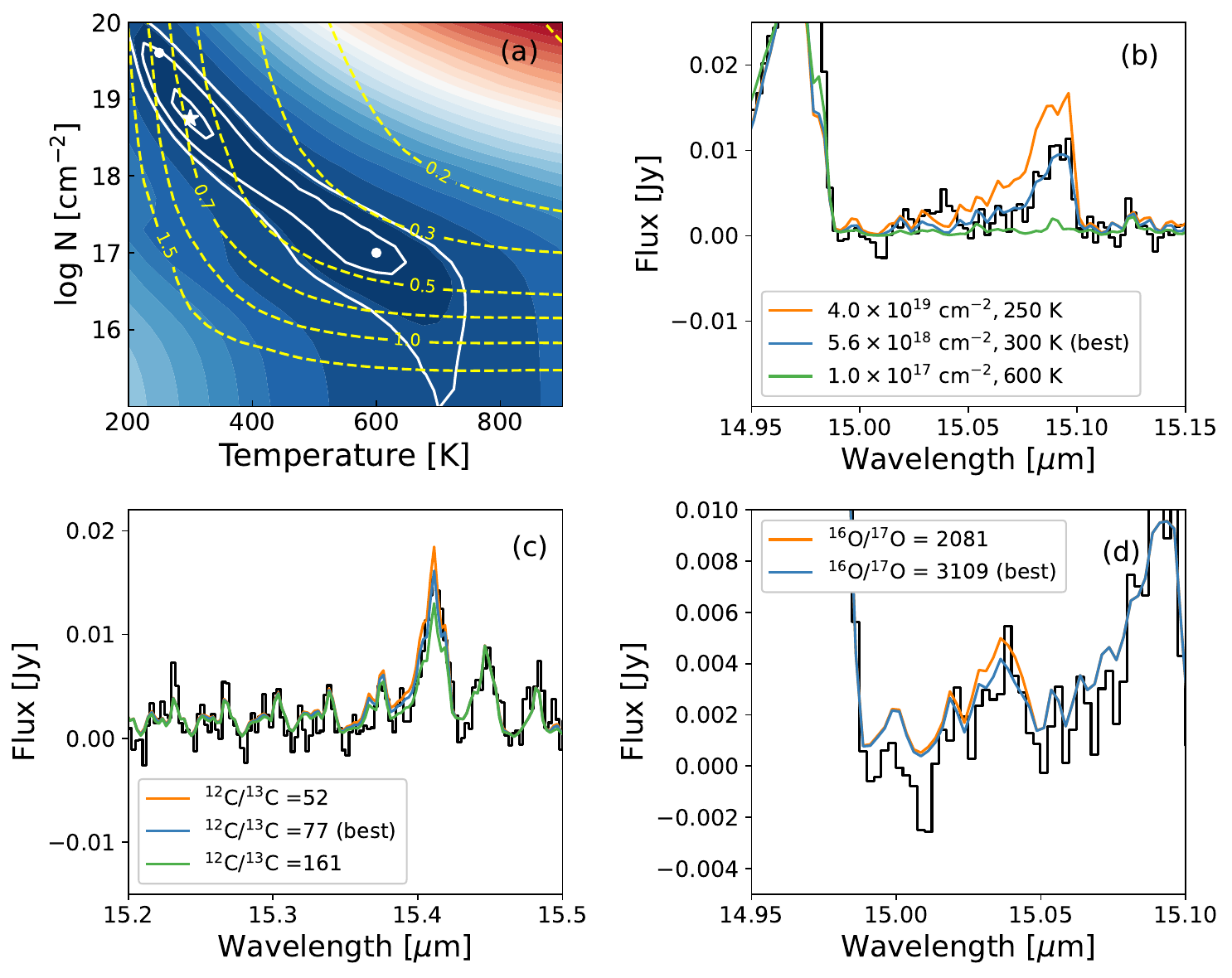}
\caption{Analysis with $^{16}$O/$^{18}$O ratio fixed to the ISM value.  (a) $\chi^2$ diagram for combined C$^{18}$O$^{16}$O and CO$_2$ fit.   The white star shows the best-fit model, while white circles show models highlighted in panel (b). White solid and yellow dashed lines have the same meaning as in Figure \ref{fig:co2_fixedCratio}(a). (b) C$^{18}$O$^{16}$O Q branch overplotted with the best-fit model and two additional models at the edges of the 2$\sigma$ $\chi^2$ contour. Legend labels show the values of $N_\mathrm{CO_2}$ and $T$  for these models. (c) $^{13}$CO$_2$ Q branch overplotted with models assuming the best-fit $^{13}$CO$_2$ ratio and those representing the best fit $\pm1\sigma$.  (d) C$^{17}$O$^{16}$O Q branch overplotted with a model with the best-fit $^{16}$O/$^{17}$O as well as one representing the best fit $-1\sigma$.\label{fig:co2_fixedOratio}}
\end{figure*}

Alternatively, we can fix the $^{16}$O/$^{18}$O to the ISM value of 557 \citep{Wilson99}, and explore the implications of this assumption.  Figure \ref{fig:co2_fixedOratio}(a) shows a $\chi^2$ diagram for a fit to the combined C$^{18}$O$^{16}$O and CO$_2$ spectrum.  The best-fit model parameters are: N$_\mathrm{CO_2}=5.6\times10^{18}$ cm$^{-2}$, T$=300\ $K and R$=0.65$ AU.  A degeneracy between N and T remains in the $\chi^2$ analysis, but Figure \ref{fig:co2_fixedOratio}(b) shows that that the C$^{18}$O$^{16}$O Q-branch feature is quite sensitive to the CO$_2$ column density.  With T and N$_\mathrm{CO_2}$ fixed to these best-fit values, we find a best-fit $^{12}$CO$_2/^{13}$CO$_2$ ratio of 77$^{+84}_{-25}$ and a 
best-fit $^{16}$O/$^{17}$O ratio $>$3109$_{-1028}$.  These fits are shown in Figure \ref{fig:co2_fixedOratio}(c) and (d).

The results of both approaches are summarized in Table \ref{table:slab_fits}.  Fits with either fixed C or fixed O isotope ratios produce similar best-fit model parameters, with T$=300-325$ K, N$_\mathrm{CO_2}=3.5-5.6\times10^{18}$ cm$^{-2}$ and R$=0.58-0.65$ AU.

\subsection{Hydrogen Cyanide}

\begin{figure*}[ht!]
\centering
\includegraphics[width=6.5in]{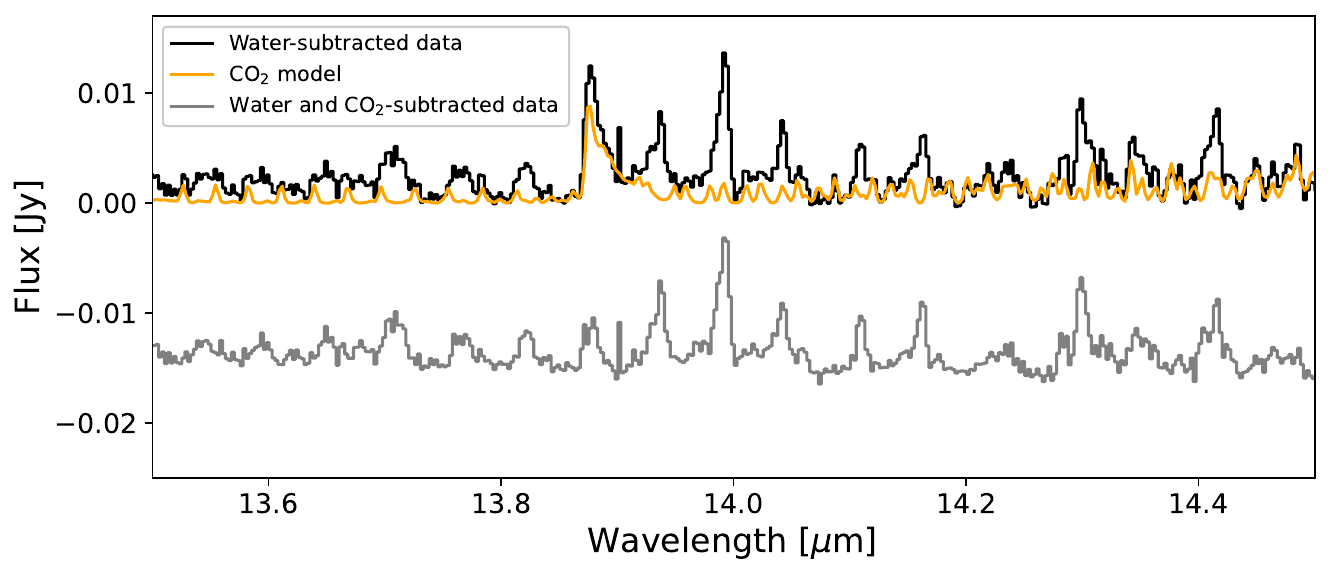}
\caption{Subtraction of CO$_2$ model from water model-subtracted data in HCN-emitting region.\label{fig:co2_subtraction}}
\end{figure*}

In order to investigate HCN emission features in MY Lup's spectrum, we first subtract both the best-fit water models and CO$_2$ models from the data.  The CO$_2$ subtraction process is shown in Figure \ref{fig:co2_subtraction}, demonstrating the substantial contribution from CO$_2$ at $\sim$13.9 $\mu$m.  The remainder of the HCN analysis utilizes the water and CO$_2$-subtracted spectrum.

\begin{figure*}[ht!]
\centering
\includegraphics[width=6.5in]{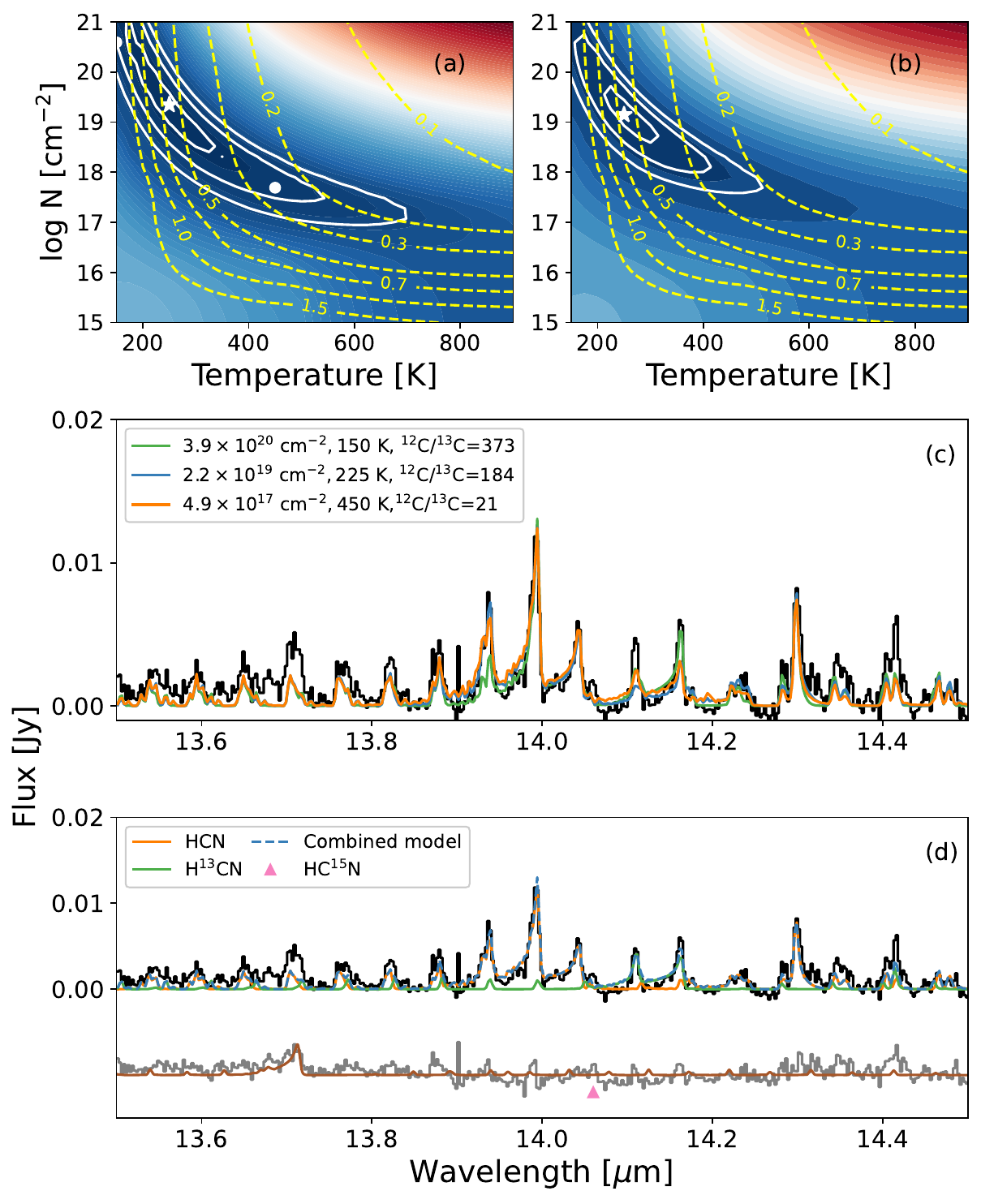}
\caption{(a) $\chi^2$ contours for a fit to HCN only.  Here and in panel (b), a white star marks the best-fit model; white circles mark additional models explored in panel (c).   White contours mark 1, 2 and 3$\sigma$ confidence intervals assuming the minimum $\chi^2_\mathrm{red}=1$. (b) $\chi^2$ contours for a  combined fit to HCN and H$^{13}$CN, with  $^{12}$C/$^{13}$C fixed to the ISM ratio. (c) Three models (with parameters marked in panel (a)) with differing $^{12}$C/$^{13}$C ratios, but only subtle changes in the quality of the model fit.  (d) Best-fit model  for  HCN and H$^{13}$CN combined ($N_\mathrm{HCN} = 1.4\times10^{19}\ \mathrm{cm}^{-2}$, $N_\mathrm{H^{13}CN} = 2.0\times10^{17}\ \mathrm{cm}^{-2}$, $T=250$ K and $R=0.61$ AU), with $^{12}$C/$^{13}$C fixed to the ISM ratio, corresponding to the star in panel (b).  A pink triangle marks a tentative detection of HC$^{15}$N, and the brown curve shows a  C$_2$H$_2$ model with T$=250$ K,  R$=0.61$ AU and N$=10^{16}$ cm$^{-2}$.\label{fig:chisq_hcn}}
\end{figure*}

Figure \ref{fig:chisq_hcn} shows the HCN-emitting region, which demonstrates evidence for both HCN and H$^{13}$CN emission. We also note an emission feature located near the HC$^{15}$N Q branch at $\sim$14.06 $\mu$m; given its tentative nature, we do not model this feature further. A fit to the HCN emission alone (excluding the region near the H$^{13}$CN feature) is performed using the same procedure as for CO$_2$; in short, we iterate through a grid of N and T to create a model using ``slabspec'', adjust the emitting area to best match the data, and compute $\chi^2$ for each model.  Figure \ref{fig:chisq_hcn}(a) shows the resulting $\chi^2$ diagram, which demonstrates a degeneracy between N and T.  In panel (c), we highlight three models within the 1$\sigma$ contour --- the best-fit model, marked by a white star in panel (a), as well as two models marked by white circles.  We additionally add emission from H$^{13}$CN assuming the same T and R as for HCN, but varying N$_\mathrm{H^{13}CN}$ (and therefore, the $^{12}$C/$^{13}$C ratio) to minimize the residuals.  We can see that these three models produce subtle variations in the resulting spectrum, and no model is clearly preferred.  Therefore, as with CO$_2$, the  C isotope ratio cannot be tightly constrained with these data.

If we instead fix $^{12}$C/$^{13}$C to the ISM ratio of 68, and simultaneously fit both the HCN and H$^{13}$CN spectral regions, assuming they arise from a reservoir with the same N, T and R, the model parameters become more tightly constrained.  The $\chi^2$ contours are shown in Figure \ref{fig:chisq_hcn}(b); the best-fit model has N$_\mathrm{HCN} = 1.4\times10^{19}\ \mathrm{cm}^{-2}$, N$_\mathrm{H^{13}CN} = 2.0\times10^{17}\ \mathrm{cm}^{-2}$, T$=250$ K and R$=0.61$ AU.  The best-fit model is also shown in Figure \ref{fig:chisq_hcn}(d). 

The model residuals show a potential feature near the C$_2$H$_2$ Q branch at $\sim13.7\,\mu$m.  If we assume the C$_2$H$_2$ originates in a reservoir with the same emitting area and temperature as for the combined HCN and H$^{13}$CN fit, we find that  N$_{C_2H_2}\lesssim10^{16}$ cm$^{-2}$, implying N$_{C_2H_2}$/N$_{HCN} \lesssim 10^{-3}$.

\section{Discussion}
\subsection{Constraints on isotopic ratios}
 The discovery of multiple isotopologues of CO$_2$ emitted by the MY Lup disk opens the possibility of studying isotopic fractionation in inner disk atmospheres.  In the context of isotope-selective photodissociation, the isotopic fractionation of a given molecule depends on whether the molecule is produced in a region affected by CO self-shielding, and also whether the molecule is produced via a chemical pathway that includes the isotopically heavy water, or isotopically light CO.  \citet{Calahan22} suggest that the disk atmosphere probed by JWST is partially affected by CO self-shielding, and that observed water may appear enhanced in $^{18}$O by a factor of $\sim 2$ relative to ISM values.  But what about for CO$_2$?  \citet{Bosman22} suggest that the observed CO$_2$ arises from a similar disk atmospheric layer as does the water, which would lead to a prediction that observed CO$_2$ should also be isotopically heavy.

In Table \ref{table:slab_fits}, we compute the measured enhancements in heavy isotopes relative to the ISM value: R$_i$/R$_\mathrm{ISM}$, where R$_i$/R$_\mathrm{ISM}>1$ implies enhancement of the heavy isotope relative to ISM values.  While CO$_2$ shows potential enhancement by a factor of 1.5 in $^{18}$O for our fixed-C ratio model sets,  $^{17}$O is depleted in those same models.   It is not clear that there is a mechanism to explain enhancement in one heavy isotope with simultaneous depletion in another.  Isotopic ratios are also consistent with ISM values to within $\sim2 \sigma$, so observed enhancements and depletions may simply reflect our uncertainties at this time.

\begin{figure*}[ht!]
\centering
\includegraphics[width=6.5in]{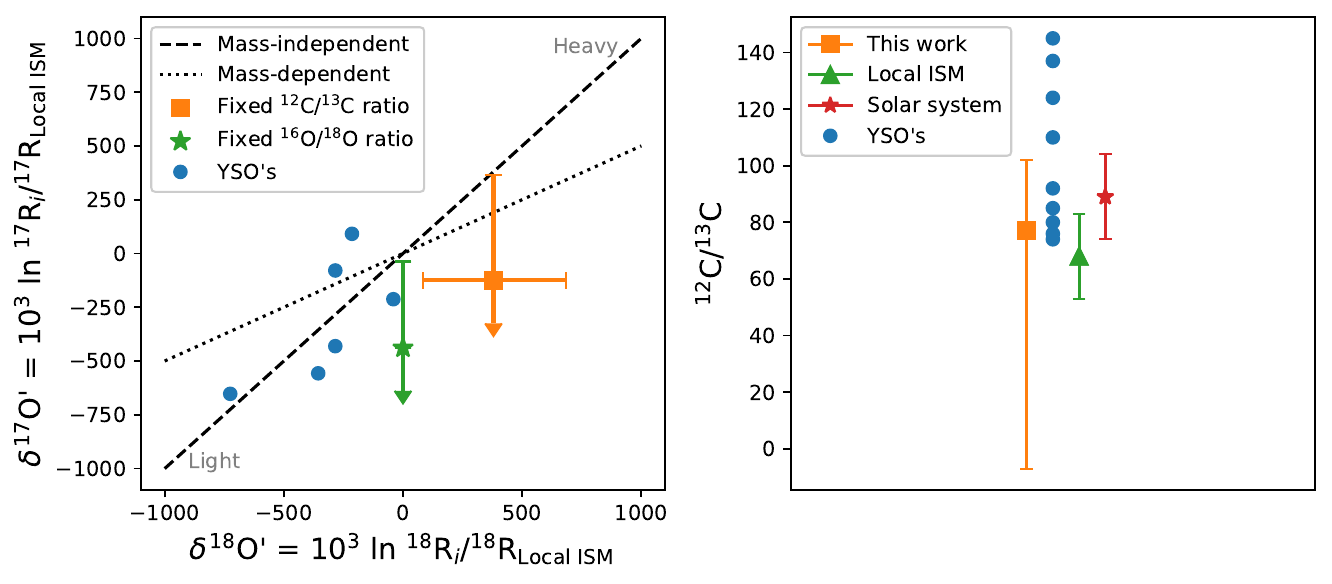}
\caption{Left: Oxygen isotope fractionation relative to ISM ratios; following standard cosmochemistry notation, heavier O is on the upper right of the plot.
 $\delta^{17}$O$'$ is defined as $10^{3}$ ln $(^{17}$R$_i$/$^{17}$R$_\mathrm{Local\ ISM})$, $\delta^{18}$O$'$ = $10^{3}$ ln $(^{18}$R$_i$/$^{18}$R$_\mathrm{Local\ ISM})$, and R$_i$ is the atomic ratio while $^{17}$R$_\mathrm{Local\ ISM}$ and $^{18}$R$_\mathrm{Local\ ISM}$ are the local ISM atomic ratios $^{18}$O/$^{16}$O and $^{17}$O/$^{16}$O, i.e., the inverse of that used in the rest of the text --- see also Table \ref{table:slab_fits}.  Dotted and dashed lines have slope 0.5 and 1, as expected for mass-independent or mass-dependent fractionation, respectively.  The green star and orange square mark our values derived from CO$_2$ emission from MY Lup, as described in Section \ref{sec:co2}.  Right: Carbon isotope ratios derived from MY Lup's CO$_2$ emission (assuming fixed Oxygen ratio, as described in Section \ref{sec:co2}), compared to values from \citet{Smith15}, the local ISM, and the solar system \citep[][and references therein]{Milam05}. Points are offset on the x axis for clarity.
\label{fig:isotopes}}
\end{figure*}

In Figure \ref{fig:isotopes}, we visually compare our derived oxygen isotope ratios from the CO$_2$ emission with the local ISM value, as well as with mass-dependent and mass-independent fractionation lines.  We also show ratios measured via CO absorption associated with young stellar objects \citep{Smith15}.  As in Table \ref{table:slab_fits}, the ratios shown here are plotted relative to the ISM ratio, such that a positive $\delta$ means an enhancement in the rarer heavy isotope.  We remind the reader that we derive our values from two different analysis approaches -- fixed C and fixed O, as described in Section \ref{sec:co2} --- and that therefore, one of the points we show was artificially fixed to the ISM value on the x axis.  We see visually that $^{17}$O is depleted according to both modeling approaches, while $^{18}$O is enhanced, although values are consistent with ISM values within the error bars. 

Figure \ref{fig:isotopes} also shows the carbon isotope ratio derived from the CO$_2$ emission, assuming a fixed Oxygen ratio, as compared to the ISM, the solar system, and the young stellar objects in \citet{Smith15}.  The carbon isotope ratio is consistent with both the ISM and solar system values, as well as some of the sources from \citet{Smith15}.

Error bars on the isotope ratios remain too large to distinguish between different fractionation scenarios, and differences in the fixed-C and fixed-O analyses suggest a similar-sized systematic error. Therefore, additional error bar refinement would be helpful to make meaningful scientific conclusion about fractionation in inner disks. In addition, our analyses all assume that the isotopologues emit from the same reservoir.  Given the complex vertical and radial temperature and density structures of protoplanetary disks, this assumption is certainly not strictly true.  Nevertheless, the detection of multiple isotopologues in this source suggests a promising avenue for future work.  Observations at higher spectral resolution could test the validity of the single reservoir assumption, and be used to better constrain emission models, although the CO$_2$ Q branches are not accessible from the ground.  As we discuss in the next section, there are also several additional disks with high reported CO$_2$ column densities that would be worthy of follow-up study.

\subsection{MY Lup's unique spectrum}

Amongst $\sim$15 published MIRI-MRS protoplanetary disk spectra (see JDISCS and MINDS program descriptions in \citealp{Pontoppidan24} and \citealp{Henning24}, respectively) and other spectra investigated thus far by the JDISCS team (34 disks total in programs 1549, 1640, 1584 and 2025; Arulanantham et al., 2025, in prep), MY Lup's spectrum is the only to show emission from C$^{18}$O$^{16}$O.  We consider here the possible causes for this unique detection.

\begin{figure*}[ht!]
\centering
\includegraphics[width=6.5in]{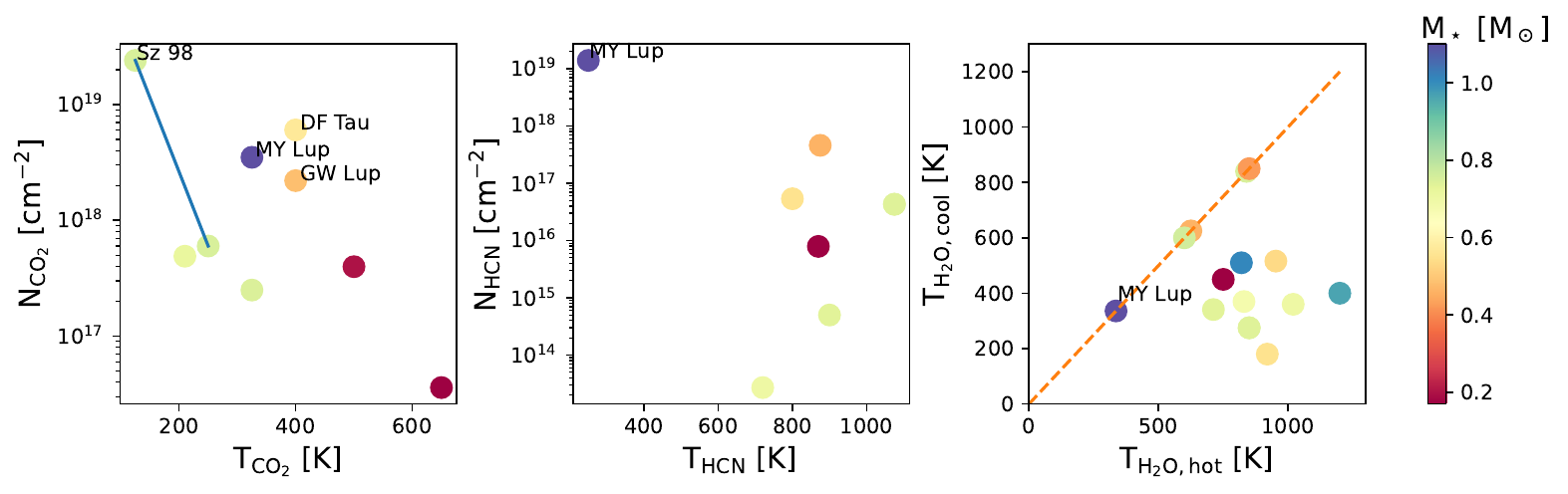}
\caption{Left: Best-fit CO$_2$ column density vs. temperature for published slab models. For 2MASS-J16053215-1933159 we show the ``Component II'' fits, which are consistent with the non-detection of $^{13}$CO$_2$ in that source \citep{Tabone23}, and for Sz 98 we show both reported models \citep{Gasman23} connected by a blue line.  Sources with the three highest column densities are labeled.  Center: Best-fit HCN column density vs. temperature.  Right: Best-fit cold vs. hot water temperatures.  Sources on the 1:1 line were fit with only a single water component.  If more than two components were used in the fit, we show the warmest and coolest components. Slab model parameters shown in all three panels collected from \citet{Gasman23,Tabone23,Schwarz24,Banzatti23,Xie23,Grant23,Pontoppidan24,Munoz-Romero24,Grant24,Perotti23,Temmink24a,Temmink24b}. Color bar shows the stellar mass in solar masses. \label{fig:demographics} }
\end{figure*}

Figure \ref{fig:demographics} shows a comparison of MY Lup's best-fit slab model parameters with other published parameters for T Tauri disks \citep{Gasman23, Tabone23, Schwarz24, Banzatti23, Xie23, Grant23, Pontoppidan24, Munoz-Romero24}.  We note that these published spectra span a wide range of stellar mass, but no particular trend emerges related to that parameter.  As discussed in Section \ref{sec:co2} and shown in Figure \ref{fig:co2_ratios}, one should expect that the detection of CO$_2$ isotopologues would be prompted by high CO$_2$ column densities, as a high column density reduces the line peak ratio between the primary and secondary/tertiary isotopologues.  MY Lup has a higher CO$_2$ column density than GW Lup, the latter of which had a reported detection of $^{13}$CO$_2$ \citep{Grant23}.  However, MY Lup has a lower reported CO$_2$ column density than Sz 98 and DF Tau.  \citet{Gasman23} report that the best-fit CO$_2$ column density for Sz 98 is uncertain, and also provide an alternative lower-$N$ model, which we also show in Figure \ref{fig:demographics}.  We find that with an ISM $^{12}$C/$^{13}$C ratio and the high-N model parameters given in \citet{Gasman23}, there should have been strong CO$_2$ isotopologue emission from Sz 98.  Therefore, we suggest that either a lower N is required, or Carbon must be highly fractionated in the Sz 98 disk. DF Tau has a water-rich spectrum \citep{Grant24}, which may have obscured the weaker CO$_2$ isotopologues.  In Appendix \ref{sec:comparison}, we compare the spectrum of MY Lup with the spectra of Sz 98, GW Lup, and DF Tau.  

The middle panel of Figure \ref{fig:demographics} also shows that the HCN column density and temperature are strikingly different for MY Lup as compared to other published disk spectra.  And the right panel demonstrates that while the water emission from MY Lup is consistent with ``cool'' water temperatures in multi-component slab model fits, MY Lup is the only disk seemingly lacking a warm/hot water component. 

\begin{figure}[ht!]
\centering
\includegraphics[width=4in]{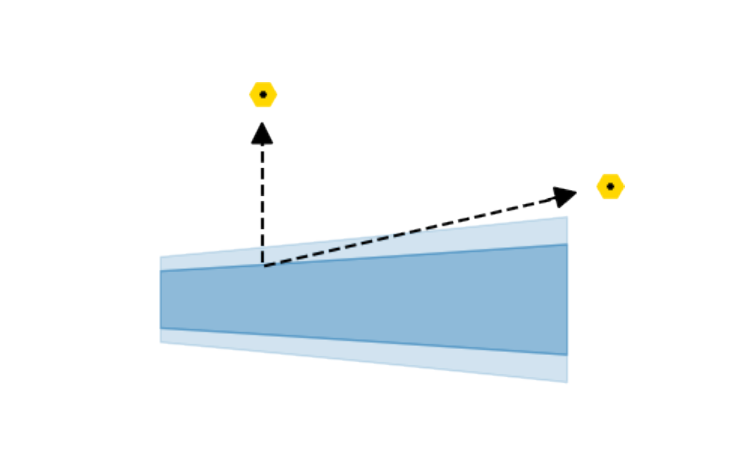}
\caption{Schematic demonstrating how an edge-on line of sight can increase the optical path through the disk atmosphere as compared to a face-on line of sight. \label{fig:los}}
\end{figure}

The presence of CO$_2$ and HCN isotopologues in the MY Lup spectrum is likely due to their high overall column densities, but why are these column densities so high in the MY Lup disk?  One distinguishing characteristic of MY Lup is its high (70-80$^\circ$) inclination.  As shown schematically in Figure \ref{fig:los}, a high inclination can reveal a larger line-of-sight column of gas above the $\tau_\mathrm{dust}=1$ surface.  For a plane-parallel slab, the path length scales as sec(i), resulting in a factor of 3--4 enhancement for MY Lup's inclination, but any dust settling ratio would increase this enhancement level.  Dust settling itself can also increase infrared molecular line/continuum ratios \citep{Meijerink09,Bosman17}, even with no inclination effect.  These possibilities seem unlikely to fully explain MY Lup's spectrum, however, as they should affect all molecular emission lines; however, MY Lup has weak water emission, only a marginal possible detection of C$_2$H$_2$, and no detections of CO or additional molecules.

Inner disk clearing likely also plays an important role in producing MY Lup's spectrum.  Although prior work has been ambivalent about the extent of clearing in MY Lup's inner disk \citep{Romero12,vanderMarel18,Alcala19}, the presence of photospheric features in the 5$\mu$m region of the MRS spectrum confirms a low degree of veiling (see Figure \ref{fig:photosphere}). Lower veiling at $\sim$5$\mu$m is consistent with reduced emission from small dust grains in the inner disk \citep{Salyk09}, as for so-called ``transition'' disks \citep{Koerner93,Calvet02}.  Also, MY Lup's molecular emission is consistent with colder temperatures as compared with other observed disks, particularly for HCN and H$_2$O (see Figure \ref{fig:demographics}).  As modeled by \citet{Vlasblom24} in an effort to explain the strong CO$_2$ emission from GW Lup's disk \citep{Grant23}, if the inner clearing extends to between the water and CO$_2$ snowlines, it may greatly reduce the water/CO$_2$ line ratios.  

\citet{Bosman17} also suggest that inward drift of CO$_2$ ice-covered pebbles could enhance inner disk CO$_2$ column densities, and predict that a cold excitation temperature would be observed for such emission.  Pebble drift in general would be expected to also enhance water vapor emission \citep[][Houge et al. 2024, in preparation]{Kalyaan21}, inconsistent with MY Lup's spectrum, but inner disk clearing may act to dissociate the liberated water vapor \citep{Bethell09}.  

It should be highlighted that although emitting areas from LTE modeling are often converted to disk radii, the conversion $R = \sqrt{A/\pi\cos{i}}$ only holds for a full disk that approximates a circular geometry.  In the case of a disk with a large inner clearing, the derived $R$ reflects the width of the emitting ring of material, but not its physical location in the disk.  Thus, we cannot say whether the CO$_2$ and/or water may arise from beyond a cleared inner region. 
 The high inclination of MY Lup's disk could also produce self-absorption in the molecular lines, which would affect measured emitted fluxes.  Kinematic information from spectrally-resolved line shapes could provide us with the physical location of the emission, and indicate the presence of any self-absorption.

The high column and low temperature of HCN (as well as the correspondingly high HCN/H$_2$O column density ratio of 14) are also unusual in MY Lup's spectrum (see Figure \ref{fig:demographics}).  While high C/O ratios could enhance HCN abundances in the disk atmosphere, high C/O would be expected to enhance C$_2$H$_2$ abundances to an even greater degree \citep{Najita11}, which we do not observe.  In addition, given the similar adsorption energies for H$_2$O and HCN \citep{Aikawa97}, one might therefore expect the inner clearing proposed by \citet{Vlasblom24} and \citet{Grant23} to also suppress HCN.  Therefore, explaining MY Lup's spectrum may require some unique chemistry that has yet to be fully explained.

All abundances in this work do assume LTE excitation, and this assumption can influence derived column densities.  For HCN, the key observational signature driving the slab model fits to high N and low T is a low relative peak ratio of the HCN Q branches (for the primary isotopologue) coupled with low flux in the high rotational states (i.e., the Q branches appear skinny, rather than fat).  Therefore, the  signature driving our model fits to low T and high N could potentially be mimicked if the vibrational states can be populated non-thermally, even while the gas temperature remains low.  Non-LTE excitation has previously been predicted for MIRI-MRS HCN lines \citep{Bruderer15}.  However, the models of \citet{Bruderer15} have neither successfully matched resolved HCN lineshapes nor the strength of vibrationally excited states \citep{Najita18}.  Since the rotational and vibrational states can be successfully fit with a single slab model, it is not obvious that non-LTE excitation is required to explain MY Lup's HCN spectrum.  The high HCN column density is also supported by the isotopologue detection.  

\section{Conclusions}
In conclusion, we report strong CO$_2$ and HCN emission, including from isotopologues of both molecules, but weak water emission from the protoplanetary disk around the T Tauri star MY Lup.  Slab modeling suggests relatively high column densities of CO$_2$ and HCN, but a typical column density of water.  Isotopologue detections are helpful in constraining column densities in otherwise degenerate modeling, although this requires the fixing of isotopologue ratios.  The temperatures of all molecules are cold compared to most other disks, particularly for HCN and water. 

Using the isotopologue detections, we investigate whether there is any evidence for isotopic fractionation.  Systematic errors and modeling assumptions limit interpretations, although the observations are consistent with ISM isotope ratios to within the error bars.  However, the detection of isotopologues in this source makes it an ideal target for follow-up observations, ideally with high resolution spectrographs.  High resolution spectra would have higher molecular line/continuum ratios, enhancing the detectability of the isotopologue signatures, and provide kinematic information to reduce modeling degeneracies.  Other protoplanetary disks with high molecular column densities would also be promising targets for such follow-up work.

Given its cold molecular temperatures, weak water emission, and high CO$_2$ and HCN column densities, MY Lup appears unique amongst protoplanetary disks observed with MIRI-MRS to date.  We attribute this unique spectrum to a combination of inner disk clearing and a high disk inclination, although additional unusual chemical conditions may also be present.

\begin{acknowledgments}
This project was supported by the STScI grant JWST-GO-01584.001-A ``A DSHARP-MIRI Treasury survey of Chemistry in Planet-forming Regions''.  A portion of this research was carried out at the Jet Propulsion Laboratory, California Institute of Technology, under a contract with the National Aeronautics and Space Administration (80NM0018D0004). This research used the SpExoDisks Database at spexodisks.com. 
\end{acknowledgments}
\clearpage

\appendix
\section{HI Emission and Absorption}
\label{sec:HI}
\begin{figure*}[ht!]
\centering
\includegraphics[width=6.5in]{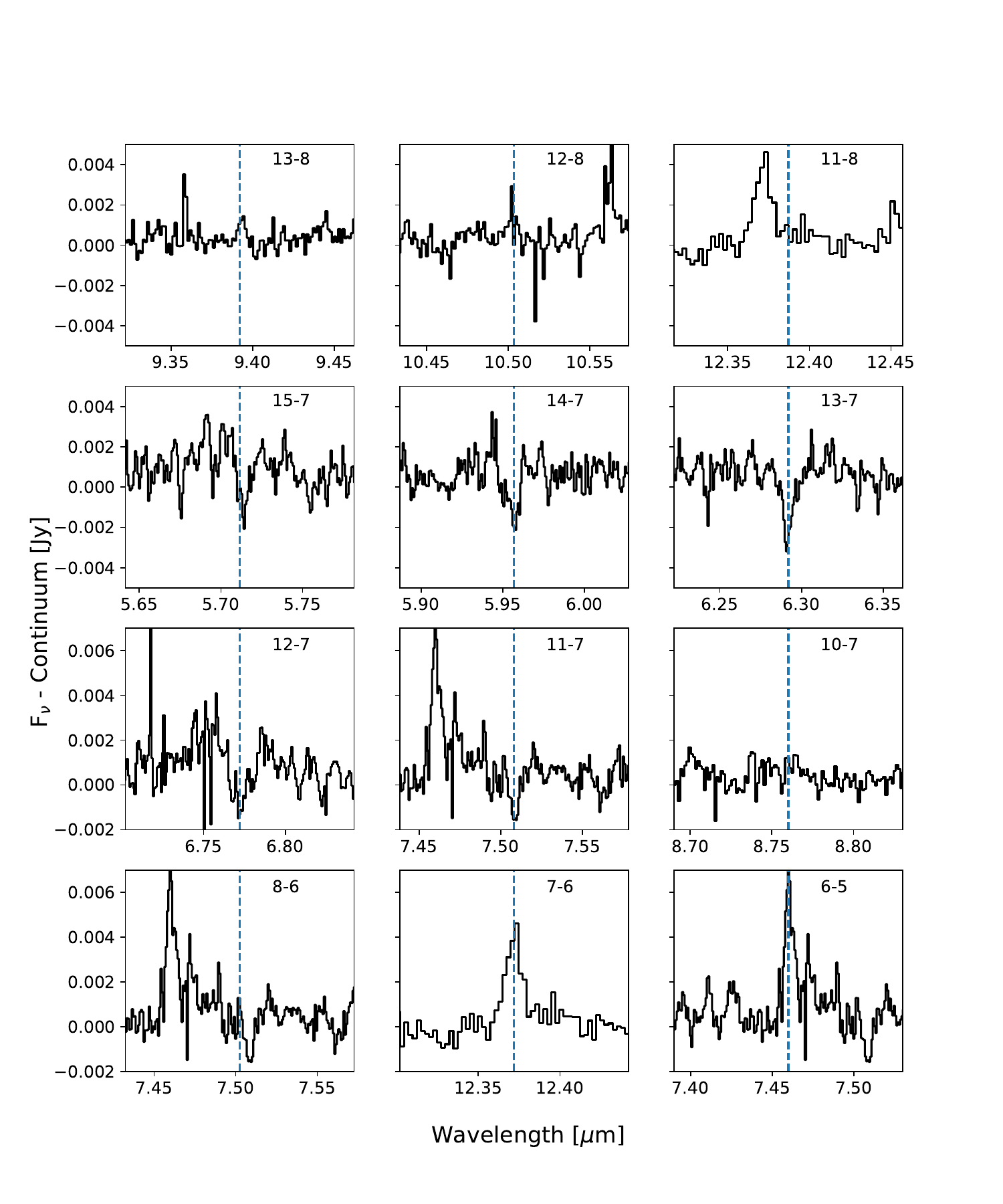}
\caption{Close-up of selected H I transitions.  Upper and lower electronic levels are shown in the upper right of each panel. \label{fig:HI}}
\end{figure*}

Figure \ref{fig:HI} shows a close up of H I transitions in the MY Lup spectrum.  The H I spectrum shows a combination of emission and absorption features. Emission from several H I transitions has been calibrated as a tracer of disk accretion \citep[e.g.][]{Muzerolle98}, including H I (7-6), which appears in MIRI-MRS data \citep{Rigliaco15}.  The presence of H I absorption could indicate absorption by foreground material in the disk or wind of this edge-on disk, although it is unclear why only some lines are in absorption (in particular, the lines with lower level $n=7$), while others are in emission.  Absorption lines all appear at the shortest wavelengths, and in MRS Channel 1, where a standard star is used as a telluric calibrator, rather than the asteroid.  However, we find no indication that the standard star is introducing spurious absorption.  It may instead be the case that we observe the stellar H I absorption at shorter wavelengths, where veiling is lower, and accretion-produced H I emission at longer wavelengths, where veiling is higher and fills in the photospheric absorption.  In-depth analysis of the H I transitions is left as future work.

\section{Evidence for non-LTE excitation of CO$_2$}
\label{sec:nonLTE}
\begin{figure*}[ht!]
\centering
\includegraphics[width=6.5in]{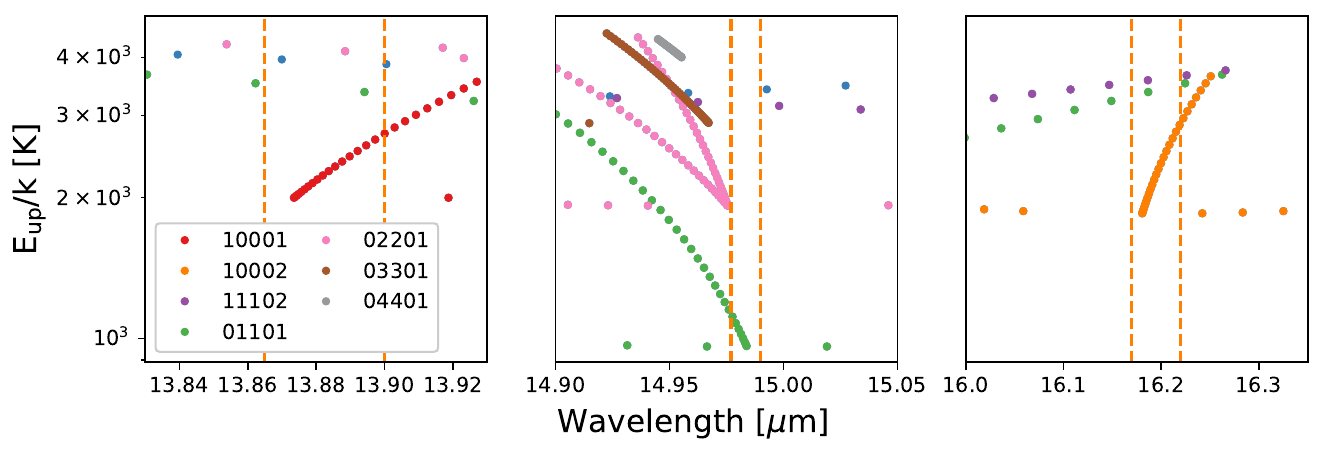}
\caption{Upper level energies of emission lines in regions highlighted in Section \ref{sec:co2} (from HITRAN database; \citealp{Gordon22}); labels show the upper state quantum numbers ($v_1$, $v_2$, $l_2$, $v_3$, $r$) where $v_i$ are ith vibrational levels, $l_2$ is the angular momentum of the $v_2$ state, and $r$ is the Fermi resonance symmetry. Dashed vertical lines mark regions with large modeling residuals discussed in the text.
\label{fig:co2_eupper}}
\end{figure*}

Figure \ref{fig:co2_eupper} shows upper level energies for the two emitting regions highlighted in Section \ref{sec:co2} and Figure \ref{fig:nonLTE}.  Labels are the quantum numbers associated with the upper level ($v_1$, $v_2$, $l_2$, $v_3$, $r$, where $v_i$ are ith vibrational levels, $l_2$ is the angular momentum of the $v_2$ state, and $r$ is the Fermi resonance symmetry), as obtained from HITRAN \citep{Gordon22} and described in \citet{Rothman05}. Dashed vertical lines highlight the regions where residuals distinguish between preferred models.  

The left- and right-hand plots highlight two features that require high column densities to match the observed spectrum.  The 13.88 $\mu$m feature is produced by a transition from $10001$ to $01101$, while the 16.2 $\mu$m feature is produced by a transition from the $10002$ to the $01101$ state.  Thus, both represent transitions between $v_1=1$ and $v_2=1$ states.

The middle plot is the primary CO$_2$ Q branch feature, and the shape of the ``red edge'' highlighted with the dashed vertical lines depends on the relative strength of the fundamental ($v'_2$=1) vs. hot-band ($v'_2\geq 2$) lines. In the context of the LTE models, low column densities are required to fit the observed ``red edge''; higher column densities create more emission in the excited states, producing a sharper dropoff on the ``red edge'' of the Q branch than is observed in the data. Since this feature depends on the states with $v'_2\geq 2$, its shape may be more affected by non-LTE excitation.

%where $v_1$, $v_2$ and $v_3$ are the vibrational quantum numbers, $l2$ is the vibrational angular momentum quantum number, and $r$ is the rank of the Fermi polyad \citep{Tashkun03,Rothman05}.

%see MYLup_co2only_chisq.ipynb

\section{Comparison with other CO$_2$-rich disks}
\begin{figure*}[ht!]
\centering
\includegraphics[width=6.5in]{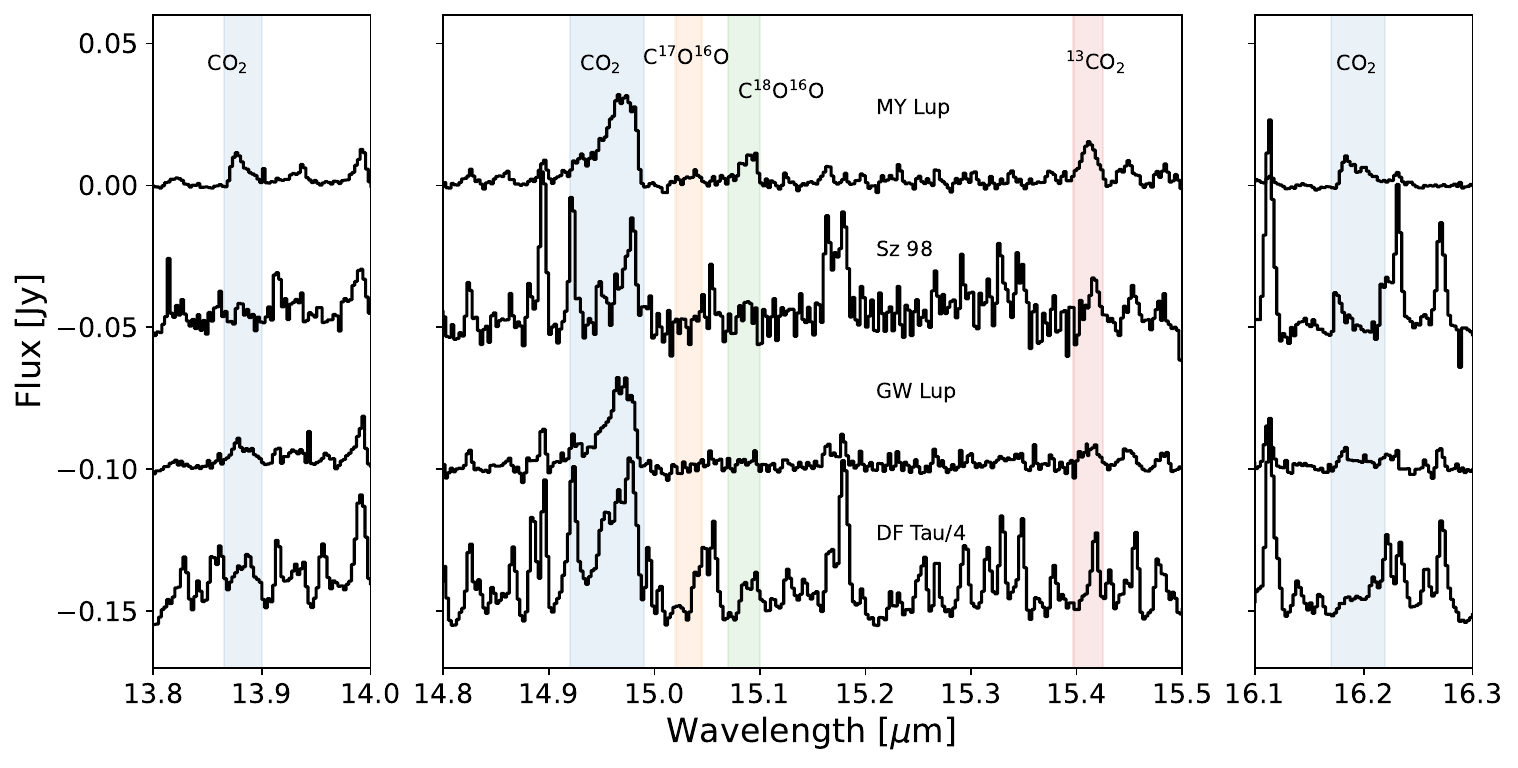}
\caption{Continuum-subtracted MIRI-MRS spectra (reduced by the JDISCS team \citealp{Pontoppidan24}) for the four sources with the highest reported CO$_2$ column densities: MY Lup (this work), Sz 98 \citep{Gasman23}, GW Lup \citep{Grant23} and DF Tau \citep{Grant24}.  The DF Tau line spectrum has been divided by four for clarity. Colored bars in the central panel mark the locations of the primary Q branches of CO$_2$ and its isotopologues.  The left and right panels show other regions with CO$_2$ emission features consistent with high N$_\mathrm{CO_2}$ 
\label{fig:source_comparison}}
\end{figure*}

Figure \ref{fig:source_comparison} shows a comparison of the CO$_2$-emitting region for the four disks with the highest reported CO$_2$ column densities.  Locations of the isotopologue Q branches are highlighted in the middle panel. The left and right panels show other regions with CO$_2$ emission features consistent with high N$_\mathrm{CO_2}$ (see Figure \ref{fig:nonLTE} and associated discussion).

\label{sec:comparison}

\begin{comment}
\subsection{Extra figures}
\begin{figure*}[ht!]
\centering
\includegraphics[width=6.5in]{mylup_h2.eps}
\caption{Portions of MIRI-MRS spectra near the H$_2$  0-0 S(1)-S(6) transitions.\label{fig:h2}}
\end{figure*}

\begin{figure*}[ht!]
\centering
\includegraphics[width=6.5in]{mylup_atomics.eps}
\caption{Atomic emission lines\label{fig:atomics}}
\end{figure*}

\begin{figure*}[ht!]
\centering
\includegraphics[width=6.5in]{mylup_water_subtraction.eps}
\caption{Water model subtraction\label{fig:water_subtraction}}
\end{figure*}
\end{comment}

\end{document}